\pdfoutput=1
\documentclass[pdflatex,sn-nature]{sn-jnl}
\usepackage[T1]{fontenc}
\usepackage{graphicx}
\usepackage{xurl} %
\usepackage{amsmath,amssymb}
\usepackage{booktabs}
\usepackage{array}

\newcommand{\placeholder}[1]{\textit{[TODO: #1]}}
\newcommand{\arxivfloat}[1]{} %
\newif\ifarxiv \arxivfalse %
\unnumbered

\arxivtrue
\renewcommand{\placeholder}[1]{}
\usepackage{etoolbox}
\usepackage{pdfpages}
\makeatletter
\patchcmd{\@maketitle}{Corresponding author(s). E-mail(s): \corrauthemail\par}{Corresponding author(s).\par}{}{\typeout{arxiv: maketitle patch not applied}}
\makeatother

\begin{document}

\title[WxCrop]{From extreme weather to agricultural consequences with agentic AI}

\author[1]{\fnm{Minghan} \sur{Yu}}
\author[2]{\fnm{Mengyu} \sur{Dai}}
\author*[3]{\fnm{Yixin} \sur{Wen}}\email{}
\author*[1,4]{\fnm{Haizhao} \sur{Yang}}\email{}
\affil[1]{\orgdiv{Department of Mathematics}, \orgname{University of Maryland, College Park},
  \orgaddress{\city{College Park}, \state{MD}, \country{USA}}}
\affil[2]{\orgname{Stony Brook University}, \orgaddress{\city{Stony Brook}, \state{NY}, \country{USA}}}
\affil[3]{\orgdiv{School of Marine and Atmospheric Sciences}, \orgname{Stony Brook University},
  \orgaddress{\city{Stony Brook}, \state{NY}, \country{USA}}}
\affil[4]{\orgdiv{Department of Computer Science}, \orgname{University of Maryland, College Park},
  \orgaddress{\city{College Park}, \state{MD}, \country{USA}}}

\abstract{Extreme weather is forecast days ahead, yet its consequences for agriculture remain hard to predict: a forecast says where a storm will strike, not which crops will suffer or how much is at risk, and the records that could tell span many sources. We present WxCrop, the first agent that judges agricultural hazards from physically interpretable weather indices examined by causal analysis, and the first that combines weather and agricultural information to predict crop loss reliably for past and forecast dates. The workflow applies to any region and hazard from that region's weather, crop and loss records. Florida is the demonstration, with hurricanes, freezes and dry heat each given an index and a loss model fitted to insurance records. Three checks show reliability: the hurricane index detects recorded cyclone days with an AUC of 0.98, the total recorded loss of held-out years is reproduced to within 2\%, and within hurricane seasons the counties in the top third of predicted citrus loss lost 48\% of their production and those in the bottom third 15\%. Users ask the agent for each crop group's loss as a fraction of expected season value, every number sourced, so that governments, growers and scientists respond to agricultural disasters quickly.}

\maketitle

\section{Introduction}\label{sec:introduction}

Weather has shaped what people can grow for as long as there has been agriculture, and it still does. Disasters cost world agriculture an estimated US\$3.26 trillion between 1991 and 2023, about 4\% of global agricultural GDP each year, with climate-related hazards accounting for close to US\$2.9 trillion of it, and warming is raising the exposure of staple crops in most producing regions even where farms adapt \cite{fao2025disasters,fao2026climateRisk,lesk2016extremes,hultgren2025adaptation,tuninetti2026drought}. Early warning is among the best-returning defences, with anticipatory action returning up to seven dollars for each one invested \cite{fao2025disasters}. Yet a warning is only as useful as what it says. A weather forecast tells us what the sky will do; it does not tell a grower, an agency or a scientist what the storm will do to the crops.

Consider a hurricane three days from landfall. A grower asks how much of this season's crop is at stake. An emergency manager asks which counties should receive inspectors and relief first. Both questions concern the same storm and the same crops, and both must be answered before landfall to be useful; today, neither can be. Florida, where we demonstrate the approach, shows the cost of waiting. Hurricane Milton struck the state on 9 October 2024 and cost its agriculture an estimated \$428 million \cite{court2025milton}, but the preliminary estimate appeared ten weeks after landfall \cite{ufifas2024miltonPrelim} and the final one about ten months after \cite{court2025milton}. The official answer arrived long after the decisions it should have informed.

The obstacle is not missing evidence but fragmented evidence. Weather reanalyses and forecasts, hydrological models, dated crop maps, growing calendars and decades of production and insured-loss records all exist, but at different scales, formats and dates, and joining them for a given field on a given day takes a specialist. Three groups pay for this. After a storm, public agencies must locate fields that may have been damaged and verify losses quickly enough to direct relief, and the yield and revenue policies of the federal crop-insurance programme on which most growers depend generally pay only after an adjuster has assessed the loss of each insured unit \cite{rmaLAM2025}, and a county's disaster designation on production losses requires a documented loss of at least 30\% of one crop \cite{cfr759}, so an early, traceable estimate of where and how much crops were damaged shortens the time to payment; the University of Florida Institute of Food and Agricultural Sciences (UF/IFAS) documented delayed planting after recent hurricanes \cite{court2025helene}, USDA identifies repair and replanting costs as reasons for disaster assistance \cite{fsa2024recovery}, and relief itself can take years, with applications for a Florida block-grant programme covering the 2023 and 2024 hurricanes opening in 2026 \cite{fdacs2026relief}. Before landfall, growers need field-specific scenarios showing which crops and how much of their current value are exposed under a forecast storm \cite{usdaFloridaHurricaneGuides}, and without them may leave otherwise productive fields idle into the next season. Scientists and extension specialists repeat the same chain of steps for every storm and every enquiry, retrieving weather records, locating parcels, matching crops and compiling exposure figures, before any judgement can be made; they need these steps carried out on request, with the evidence behind each number kept visible, so that their time goes to the decision rather than the data handling.

Existing tools look backward. The formal UF/IFAS estimates rest on the Hurricane Composite Intensity Index, which scores observed wind, rainfall and flood depth on agricultural parcels, sums the three scores, and assigns each level a commodity loss rate drawn from producer surveys and past storms \cite{court2025milton,ufifas2024miltonPrelim}; it is computed after the storm, and its additive form cannot represent hazards that compound, although interacting drivers often cause the largest impacts \cite{zscheischler2018compound}. Two earlier lines of work give a foundation for an index that compounds. Chen et al.\ combined a cyclone's wind and rainfall into one index for the whole event, which reproduced recorded typhoon damage in China \cite{chen2019ipwt}, and field records from Taiwan show that wind coupled with heavy rain, successive events and the length of a crop's production period govern crop losses \cite{lai2017taiwan}. Neither is computed by day and county or joined to the crops beneath it. Remote sensing is faster but still post-event: USDA overlays satellite flood extent and observed wind and rainfall on the Cropland Data Layer \cite{nass2024miltonAnalysis}, a UF/IFAS assistant answers growers' damage questions from Sentinel-2 imagery \cite{tziolas2026geogpt,wusf2025gaia}, and field-scale damage has been estimated from modelled typhoon winds and satellite rainfall \cite{blanc2016typhoon}; imagery needs an overpass and, for optical sensors, clear skies. Empirical studies quantify how heat, drought and compound extremes reduce production at regional to global scales \cite{hultgren2025adaptation,tuninetti2026drought,kabtih2025hotpluvial}, and recent work uses USDA Risk Management Agency (RMA) records of insured losses as a proxy for weather-driven crop damage \cite{reyes2019croploss,yao2026drought,lazin2025flood}, but at coarse resolution over seasons or years. Where forecast-driven impact models exist, they estimate housing damage \cite{kooshki2024ibf} or rank counties by population exposure. In parallel, language-model assistants are also addressing geoscience and agricultural problems conversationally: ClimSight routes questions to climate-model output \cite{kuznetsov2025climsight}, WildfireGPT coordinates retrieval agents for wildfire adaptation \cite{xie2025marsha}, DisasTeller chains vision-language agents through post-disaster assessment and notes that early errors propagate downstream \cite{chen2026disasteller}, language models have extracted flood records from hazard narratives \cite{zhou2025flashflood}, and an assistant integrates NASA tools with agentic search over NASA datasets to provide geoscience information and solve problems \cite{yu2026agentic}. These systems answer from documents, projections, imagery or data catalogues rather than from dated agricultural parcels, and none links a hazard signal on a given day to the crops and fields beneath it. Their adoption also raises trust: users doubt the sources, cannot follow the computation, and fear fabricated answers, a fear grounded in the documented tendency of language models to produce fluent but unsupported content \cite{ji2023hallucination,williams2026trustworthy,reichstein2025earlywarning}. A tool that is to be adopted must answer all three.

Taken together, these needs define what an assistant for agricultural hazard assessment has to provide. It must be usable by agencies, growers and specialists alike without technical preparation, so that plain-language requests replace manual data assembly. It must return evidence for any date a user asks about, past or forecast, rather than for a few pre-analysed events. It must apply the same workflow to every storm instead of being rebuilt for each one. And every figure it reports must be traceable to the records it was computed from. To our knowledge, WxCrop is the first workflow to meet these requirements together. In particular, it provides the first county-level weather hazard index for agriculture whose effect on crop loss has been examined by causal analysis and confirmed against records outside the insurance data. It is also the first system that combines weather and agricultural records into reliable estimates of the percentage of crop value lost, both after a hazard and ahead of it.

We present WxCrop (weather $\times$ crop), an interactive assistant that connects county-level hazard signals with the agricultural evidence behind them, built to meet these requirements. It joins public weather records and forecasts with dated crop maps, growing calendars and insured-loss records, so that every analysis can run for a past storm or for one still offshore. For a chosen date and hazard, the map shows the weather inputs that drive that hazard and a weather-only index for it, alongside the crops, their growth stage and vulnerability. The assistant answers each user's question with a dedicated product: for decision makers, a hazard index computed from the weather alone that places a county in a tier anchored in how often such weather has been followed by loss; for growers, the fraction of each crop's expected season value likely to be lost and the probability that a loss occurs at all; for scientists, the harmonised records behind both. Each answer is built to be trusted: every input is a named public record, the index decomposes into its terms, the percentages come from a fitted model rather than from the language model, and the conversational agent retrieves and explains these figures rather than generating them. We demonstrate the assistant through a case study of Hurricane Ian that follows the three steps a user takes, from the weather and crops on the map, to the hazard index, to the loss estimate, and we then check the answers against recorded losses.

WxCrop is a general procedure, and Florida is its demonstration. Each step takes a kind of record that agricultural regions around the world hold, a weather archive and forecast, a dated crop map and a record of losses, so applying it to another region or hazard means replacing the sources and refitting the model, with the way the indices are built, tested and checked unchanged. We chose Florida because its hurricanes, freezes and dry heat are costly and well documented. The longer aim is an open, traceable early-warning layer for weather impacts on crops, one that shortens the time between a storm and the resources that follow it.

\section{Results}\label{sec:results}
\begin{figure}[tbp]
  \centering
  \includegraphics[width=\linewidth]{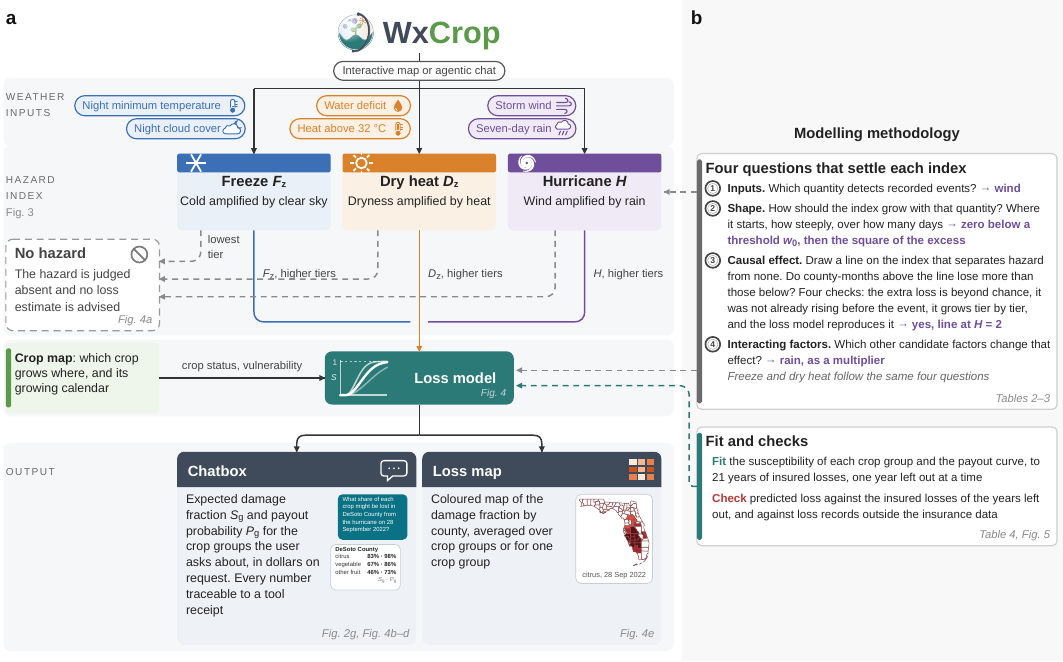}
  \caption{\textbf{Overview of WxCrop.}
    \textbf{a}, What a user sees. A question in the chat or a click on the map selects a hazard, a county and a date. Two weather layers per hazard are combined into a weather-only index, $F_z$, $D_z$ or $H$. A county in the lowest tier is judged free of the hazard and receives no loss estimate. Otherwise the index enters the loss model with the crops of the county, their growth stage and their vulnerability on that day, and the model returns for each crop group the expected loss fraction $S_g$ and the payout probability $P_g$, in the chat and as a county map. The thumbnails are real outputs for Hurricane Ian in DeSoto County on 28 September 2022. \textbf{b}, How the indices and the loss model were built: the four questions of Methods, with the hurricane answers, and the fit and checks of the loss model. Figure and table numbers in the panels point to the evidence for each step.}
  \label{fig:overview}
\end{figure}

WxCrop resolves a hazard event in three steps (Fig.~\ref{fig:overview}a): it reads the weather inputs of the hazard and the crops beneath them (Fig.~\ref{fig:workflow-platform}), computes the hazard index (Fig.~\ref{fig:workflow-hazard}), and runs the loss model, whose outputs are a reply in the chat and a county map (Fig.~\ref{fig:workflow-loss}). Fig.~\ref{fig:overview}b shows how the indices and the loss model were built and checked (Fig.~\ref{fig:trust}). We follow Hurricane Ian, which made landfall in Florida on 28 September 2022, in DeSoto County, and take a freeze and a drought with heat through the same steps. Datasets are named by their abbreviations, which Table~\ref{tab:data} spells out.

\subsection{Weather layers, crop layers and vulnerability on one map}
\begin{figure}[tbp]
  \centering
  \includegraphics[width=\linewidth]{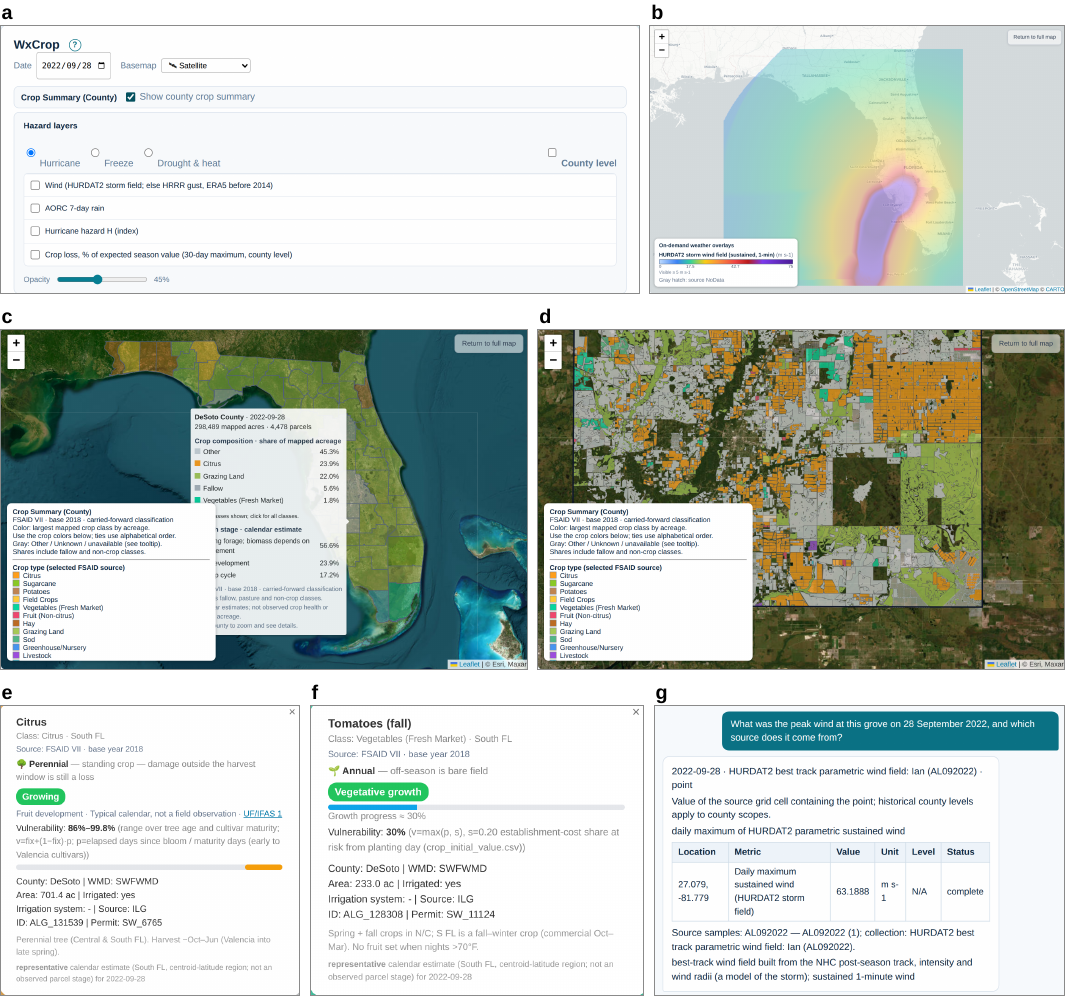}
  \caption{\textbf{Weather and crop state on one day in WxCrop.}
    Screenshots of the application for Florida on 28 September 2022, the day Hurricane Ian made landfall. \textbf{a}, Date and hazard controls. \textbf{b}, Wind layer, one of the two hurricane weather layers: on storm days the best-track wind field of Ian from NHC HURDAT2 (sustained wind), on other days HRRR gusts. \textbf{c}, County crop summary, hovering over DeSoto County. \textbf{d}, DeSoto County parcels coloured by crop class. Panels c and d use version VII of the Florida Statewide Agricultural Irrigation Demand (FSAID) inventory, classified for 2018 and carried forward. \textbf{e,f}, Popups for a citrus grove (ALG\_137275) and a tomato field (ALG\_128245), each with its growth stage and vulnerability. \textbf{g}, A question about the grove's peak wind and the reply, a modelled best-track value for the containing cell.}
  \label{fig:workflow-platform}
\end{figure}

In its first step, WxCrop lets the user choose a hazard in a panel (Fig.~\ref{fig:workflow-platform}a). The panel then offers the weather layers that drive that hazard, selected by causal analysis of insured losses (next section). For the hurricane these are wind and seven-day rainfall. Ticking a layer and a date, here 28 September 2022, loads it on demand from public archives, on its native grid or as county values, reconstructing the conditions of that day. For wind, WxCrop switches source with the weather. On days a tropical cyclone puts at least 10\,m\,s$^{-1}$ on Florida, the layer is the storm's wind field reconstructed from the National Hurricane Center best track (HURDAT2, Fig.~\ref{fig:workflow-platform}b), because gridded analyses and forecasts smooth the cores of tropical cyclones and underestimate their winds. On other days it is the daily maximum gust from analyses. The same best-track field feeds the hazard index, and the assistant flags forecast winds as such. Layers cover 2004 to the present and, using blended forecast guidance, the coming week. Each layer states its source, its grid and whether its values are observed, analysed or forecast, so that every colour can be traced back to a record. The colour boundaries of each layer are percentiles of its own historical values, from the median to the 99th, so that a colour means the same degree of unusualness on any date, with storm-class boundaries for the best-track wind field (Methods).

Beneath the weather sit the crops. The county summary colours each county by its largest crop class, and hovering over DeSoto County (Fig.~\ref{fig:workflow-platform}c) reports 298{,}489 mapped acres in 4{,}478 parcels, of which 23.9\% are citrus, 22.0\% grazing land and 1.8\% fresh-market vegetables, together with the mix of calendar growth stages. Zooming in to the county interior loads the parcel view (Fig.~\ref{fig:workflow-platform}d), drawn from more than 170{,}000 polygons in the state's agricultural land inventory, each with its crop class, specific crop and source version, as the legend states.

Clicking a parcel (Fig.~\ref{fig:workflow-platform}e,f) opens a popup with the parcel's crop, its representative growth stage on the selected date from a growing calendar, and a vulnerability percentage: the fraction of the parcel's expected season value that has accrued, and is therefore exposed, on that date. For an annual crop it is the seasonal progress, floored at the establishment cost already sunk on planting day. The tomato field in Fig.~\ref{fig:workflow-platform}f, about 27 of 90 days after its calendar planting date, stands at 30\%. For citrus the fruit progress is computed in the same way, the days elapsed since bloom divided by the days to maturity, bounded by the fastest and slowest cultivars, and added to the standing value of the trees, so that the grove in Fig.~\ref{fig:workflow-platform}e shows a range, 86\% to 99.8\%, which stays below the full crop until harvest begins. The rules for each crop type are in the code repository (Code availability). Vulnerability has two parts: the cost already invested, which is the value of the trees for perennial crops and the establishment cost for annual crops once planted, and the calendar fraction of the season's expected income, an opportunity cost that grows toward harvest. A citrus grove therefore carries at least the value of its trees throughout the year, including outside the harvest season, and a cotton field in November is mostly harvested and contributes little. This vulnerability percentage is a sourced indicator, built on growing calendars compiled from UF/IFAS crop guides and USDA statistics, of how vulnerable each parcel is on the selected date.

Everything on the map can also be asked about in plain language, at the scale the user needs: a single parcel, the current view, a county or a named place. With the grove selected, the question of its peak wind on 28 September 2022 and the source of that number (Fig.~\ref{fig:workflow-platform}g) is answered by a weather tool: 63.2\,m\,s$^{-1}$, the daily maximum sustained wind of the Ian best-track field in the grid cell (about 2\,km) containing the grove, with its source and units and a note that it is a modelled value for the grid cell. A user can equally ask which crops dominate a county, how much citrus has been lost to decline, what growth stage cotton has reached on a given date, or how much rain fell on a grove in the previous seven days. The assistant retrieves every number through defined tools, each of which returns the data version, filters and units behind the value, so that the figures in a reply are computed by WxCrop and traceable to their records rather than written by the language model. The assistant also makes the map convenient to operate: it selects the hazard from the question, moves the map, sets the date and opens a county, so that a question is enough to reach the answer.

Besides the hurricane, WxCrop covers two other major hazards for Florida agriculture, freezes and drought with heat (Fig.~\ref{fig:workflow-platform}a). Choosing either one offers its own weather layers, which work in the same way. Weather, crop, growth stage and vulnerability are therefore all on one map for each hazard, and a user can move from a statewide view to a single field without leaving the conversation.

\subsection{Physically based, causally tested compound weather indices}
From the weather layers of each hazard alone, WxCrop computes a composite hazard index that shows how unusual the weather was and rises with the damage it does, for any county on any day for which the weather inputs exist, and adds it to the map as a new layer. For hurricanes this is the hurricane hazard index, $H$, and its design follows the physics of the damage. Strong wind together with heavy rain loads wet soil and plants more than either alone, because trees in saturated soil lose much of their resistance to overturning \cite{defossez2021saturation}, and drivers that act together often cause the largest impacts \cite{zscheischler2018compound}. To find which factors matter for $H$, and which physically plausible interactions between them are present, we examined the commonly used weather quantities linked to hurricane damage, among them wind speed and duration, rainfall, ponding and wet hours, water tables, soil drainage, the interval since the previous storm and tree age. Each was tested against insured losses by causal analysis, with the candidates and the acceptance rules fixed before the results were seen. Wind and rainfall mattered, and so did their interaction, while the other candidates repeated these two, showed no signal or lacked data (Methods). $H$ therefore multiplies a wind load, which grows with the square of the sustained wind as dynamic pressure does, by an amplifier that grows with rainfall (equation~\ref{eq:hazard}). Of eight candidate forms, this one matched recorded storm days and the post-storm loss reports best.

\begin{figure}[tbp]
  \centering
  \includegraphics[width=\linewidth]{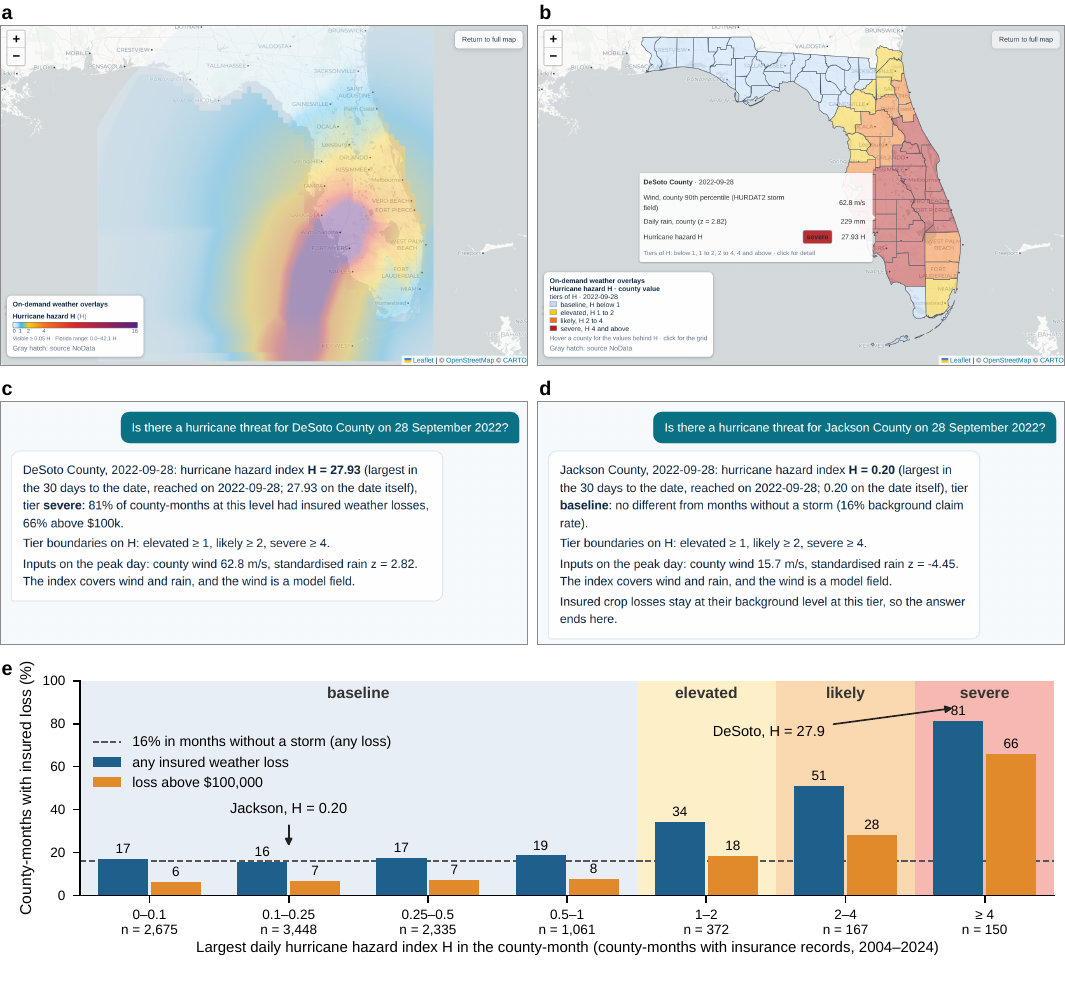}
  \caption{\textbf{One number for the hazard.}
    All panels are for 28 September 2022. \textbf{a}, The hurricane hazard index $H$ on its native grid. \textbf{b}, The county-level layer for the same index, with the hover card of DeSoto County. \textbf{c,d}, The assistant's answers for DeSoto County, near the storm's track (severe tier), and for Jackson County, in the Panhandle (baseline tier). \textbf{e}, The proportion of county-months with insured weather losses (blue) and with losses above \$100{,}000 (orange) at each level of $H$, with the four tiers shaded.}
  \label{fig:workflow-hazard}
\end{figure}

On the case day the hurricane index is highest along the path Ian took (Fig.~\ref{fig:workflow-hazard}a). Its colour bar marks the boundaries of four tiers, baseline below 1, elevated from 1 to 2, likely from 2 to 4 and severe from 4 upwards, and the county-level layer shows the same index as one value per county, coloured by tier, with a hover card that gives the county's wind, rainfall and $H$ with its tier, here for DeSoto County (Fig.~\ref{fig:workflow-hazard}b).

Asked whether DeSoto County is under threat, the assistant returns $H = 27.9$ and places it in the severe tier (Fig.~\ref{fig:workflow-hazard}c), and asked the same of Jackson County, in the Panhandle, it returns $H = 0.20$ and the baseline tier (Fig.~\ref{fig:workflow-hazard}d). Each answer gives the index, its tier and the county's wind and rainfall, and states that the wind is a model field and that $H$ covers wind and rain. The county wind is the 90th percentile, over the county's cells, of the storm's peak best-track wind, so it differs from the value at a single point such as the grove in Fig.~\ref{fig:workflow-platform}g.

Insured losses rise with $H$ (Fig.~\ref{fig:workflow-hazard}e, which groups the county-months with insurance records in 2004--2024 by the largest daily $H$ in the month, shows for each group the proportion with insured weather losses and with losses above \$100{,}000, and marks where the two counties fall). For the hurricane index, county-months in the lowest tier reported insured crop losses about as often as months with no storm at all, 16\%, while in the highest tier 81\% of county-months had insured losses and 66\% had losses above \$100{,}000. This shows that $H$ orders county-months by their insured losses. The tier boundaries at 1, 2 and 4 are the median, upper quartile and top decile of $H$ among the county-months with winds above 17.5\,m\,s$^{-1}$, the wind speed at the outermost radius reported in the HURDAT2 best-track archive, rounded and the same for every county and date, and the insured-loss frequencies are reported on the same boundaries. The assistant states this basis alongside the number, so that a user sees how often weather of this kind has been followed by loss.

Freezes and dry heat follow the same design. Each hazard has its own weather layers and an index built from them in the same way. For freezes the layers are the night's minimum temperature and cloud cover, and the index is the depth of the minimum below each farmland cell's design temperature, the temperature that 95\% of its winter nights stay above, amplified under clear skies, which govern radiative cooling. For dry heat the layers are a standardised 120-day deficit of precipitation relative to evapotranspiration and the heat accumulated above 32\,$^\circ$C over the preceding 30 days, and the index is the deficit amplified by that heat. The freeze index is computed for each night, and the dry-heat index accumulates over the growing season. Their inputs were selected by the same protocol (Methods). Their tier boundaries are fixed thresholds set in the same way as those of $H$, from the exposed records: for freezes the median, 90th and 97th percentiles of the county-months in which the night fell below the design temperature, and for dry heat the median, upper quartile and top decile of insured field-crop county-seasons.

\subsection{Fitted expected-loss percentages for each crop group}
\begin{figure}[tbp]
  \centering
  \includegraphics[width=\linewidth]{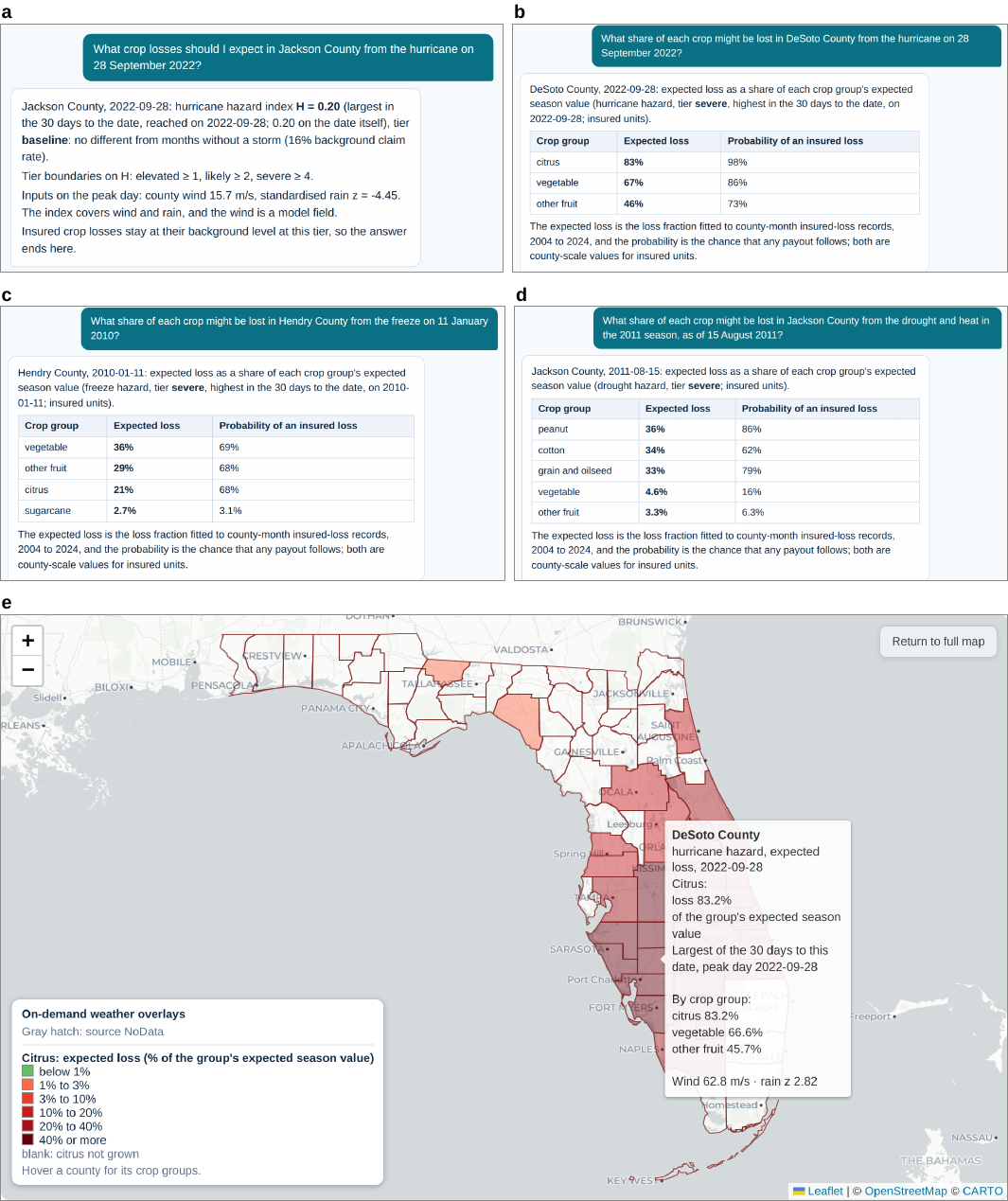}
  \caption{\textbf{From hazard to a percentage of the crop.}
    \textbf{a}, A county in the lowest hazard tier (Jackson County, hurricane, 28 September 2022): the answer ends at the tier. \textbf{b}, A county in the severe tier on the same day (DeSoto County): expected loss by crop group. \textbf{c,d}, The same output for a freeze (Hendry County, 11 January 2010) and for dry heat (Jackson County, 15 August 2011). \textbf{e}, The county-level layer for citrus on the same day, with the hover card of DeSoto County.}
  \label{fig:workflow-loss}
\end{figure}

The hazard tier decides what the assistant does next. If the index falls in the lowest tier, as it does for Jackson County on the case day, the assistant tells the user that the day matches one without that hazard and that insured crop losses stay at their background level; the answer ends at the tier, because for such days an estimate would sit at the background and lend false precision to a non-event (Fig.~\ref{fig:workflow-loss}a). If the index is higher, the assistant treats the day as a disaster of that kind and takes everything from the first step as inputs: which crops are present in the county, their growth stage on that day, and their vulnerability. These enter the loss model fitted for that hazard, which returns the expected loss of each crop group as a percentage of its expected season value.

How strongly a hazard affects a crop depends mainly on two factors. One is the susceptibility $\theta_g$ of crop group $g$, how sensitive that crop is to the kind of hazard, which differs between crops and between hazards: wind strikes cotton hard and sugarcane hardly at all, and drought hits peanuts while irrigated citrus responds little. The other is its yield vulnerability on the day, $\varphi_g$, the fraction of the season's yield that is in the field. It is the vulnerability above without its fixed part, the value of the trees, which lies outside what crop insurance pays for and is unrelated to the percentage loss of the season's yield that we aim to predict. Citrus stands in the field all year, whereas vegetables in late September have only part of their season in the field, so the same storm costs them less. For DeSoto County, in the severe tier, the loss model combines them with the index into two quantities for each crop group (Fig.~\ref{fig:workflow-loss}b): the expected loss fraction $S_g$, which rises with the index, $\theta_g$ and $\varphi_g$, and the probability $P_g$ that insured growers of that group receive any payout, the insurer's payment for a loss, in the month, because many county-months with moderate hazard end without a payout. The assistant reports both. They refer to the insured units of the crop group, the scale on which the model was fitted (Methods). The index is taken as its largest daily value in the 30 days that end on the selected date (the season to date for dry heat), which follows the insured losses used to fit the model: they are recorded by month and, once incurred, stay in the monthly record, so that the estimate holds at its peak as the storm passes instead of falling back day by day. For citrus $S_g$ is 83\% of the season's expected value and $P_g$ 98\%. Percentages are county-level values for each crop group, which match the scale of the insured-loss records used to fit them.

Similarly, when asked about a county and a date at which the index of another hazard is high, the assistant returns the corresponding expected-loss table for every crop group the model covers (Fig.~\ref{fig:workflow-loss}c,d), here for the freeze of 11 January 2010 in Hendry County and for dry heat in Jackson County on 15 August 2011, both in the severe tier.

The same loss model also draws a map. The fourth layer in the hazard panel colours every county by the expected loss to its crops, as a percentage of their expected season value, and lists the crop groups behind each value on hover. A drop-down shows one crop group at a time, with the counties that do not grow it left blank, here citrus (Fig.~\ref{fig:workflow-loss}e). The layer is computed on demand, either by asking the assistant about a date or with a compute button, and appears for dates whose values are cached. For 28 September 2022 the citrus counties along the track of Ian in the south-west have the highest values.

The three steps keep the conversation proportionate to the evidence. A user who asks about a quiet day receives a map and a short reassurance. A user who asks about a hurricane, a freeze or a dry season receives the crops at risk, how much of their value is likely to be lost, how likely that loss is, and where each number comes from. In both cases the path from weather record to answer is visible in the session, which is what makes the answer usable by people who must act on it and defend it.

\subsection{How far the answers can be trusted}
\begin{figure}[tbp]
  \centering
  \includegraphics[width=0.86\linewidth]{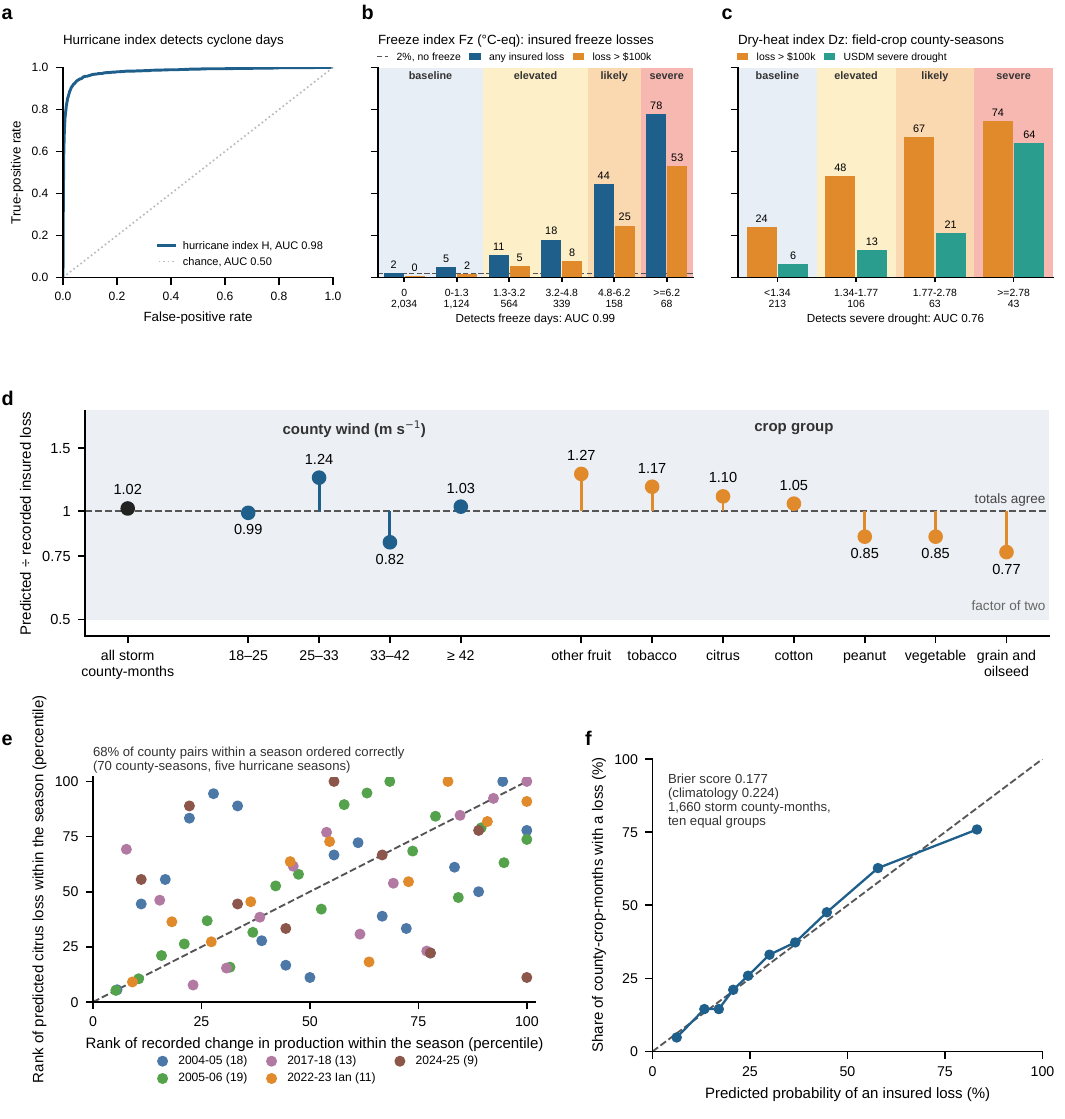}
  \caption{\textbf{How far the answers can be trusted.}
    \textbf{a}, Receiver operating characteristic curve of the hurricane index $H$ against recorded tropical-cyclone days. \textbf{b}, Proportion of winter county-months with insured freeze losses (blue) and with losses above \$100,000 (orange) in each range of the freeze index, with the number of county-months below the axis. \textbf{c}, Proportion of insured field-crop county-seasons with a claim above \$100,000 (orange) and of seasons with severe drought in the U.S. Drought Monitor (green) in each tier of the dry-heat index. \textbf{d}, Hurricane predictions against recorded insured losses, summed over storm county-months with each year held out. 1 means the totals agree and the grey band is a factor of two. \textbf{e}, Rank of the predicted citrus loss against the rank of the recorded change in citrus production within each of the five hurricane seasons, 70 county-seasons coloured by season. \textbf{f}, Calibration of the probability of an insured loss, in ten equal groups of storm county-months.}
  \label{fig:trust}
\end{figure}

The hazard indices and percentage losses WxCrop returns are advisory estimates, meant to show a user how severe a hazard is likely to be for each crop group and where to look first, and this section shows their validity at that level. Such a test needs loss labels that can be tied to the weather that caused them, and we assembled these ourselves: insured-loss records separated by the recorded cause of loss at the scale of county, crop group and month, with records outside the insurance data alongside. Against these we will show two things in turn: that the three hazard indices detect and order the hazards they stand for, and that the loss model reproduces the insured losses it was fitted to, agrees in ranking with records from outside the insurance data and returns a calibrated probability of a payout (Fig.~\ref{fig:trust}).

First, to show that the hazard indices detect the hazards they stand for, we use the area under the receiver operating characteristic curve (AUC) \cite{hanley1982auc}, the probability that a randomly chosen day with a recorded event receives a higher index than a randomly chosen day without one, which is 0.5 for a random score and 1 for a perfect one. These reference records contain no crop information. The hurricane index $H$ separates recorded tropical-cyclone days from other days with an AUC of 0.98 (Fig.~\ref{fig:trust}a), and the test of its ordering was given above, where insured losses rise with its tier (Fig.~\ref{fig:workflow-hazard}e). The freeze index separates recorded freeze days from other days with an AUC of 0.99, and insured freeze losses rise with its tier, from 2\% of winter county-months without a freeze to 78\% in the highest tier, with 53\% above \$100{,}000 (Fig.~\ref{fig:trust}b). The dry-heat index separates seasons with severe drought in the U.S. Drought Monitor with an AUC of 0.76 for the season to date, which looks modest as a number but is good agreement with a reference that blends several drought indicators and follows slow deficits more closely than crop losses (Methods), and across its tiers the proportion of insured field-crop county-seasons with a payout above \$100{,}000 rises from 24\% to 74\% and the proportion of seasons with severe drought from 6\% to 64\% (Fig.~\ref{fig:trust}c).

Second, to show that the loss model reproduces the insured losses it was fitted to, we predicted each year with a model whose susceptibilities, fitted on the other years of two decades of county-level insured-loss records separated by the recorded cause of loss, had never seen that year. The predicted loss summed over storm county-months matched the recorded total to within 2\% (Fig.~\ref{fig:trust}d). Resampling the years gives the ratio of the two totals a 95\% interval of 0.86 to 1.24, and in each of the ten years that hold at least 2\% of the recorded loss the prediction for the year left out was 0.65 to 1.52 times the record (Supplementary Table~8).

The county order that the model gives also agrees with records outside the insurance data, which include uninsured growers. We score it on pairs of counties within one season. A pair is ordered correctly when the county with the larger predicted loss also lost more in the record. A ranking unrelated to the record scores 50\% and a perfect one 100\%. Against county citrus production from the National Agricultural Statistics Service (NASS), in the five seasons in which county wind reached hurricane force in a citrus county, the model orders 68\% of 493 pairs correctly (95\% interval 59 to 76\%, from resampling the counties) and 82\% in the Ian season (Fig.~\ref{fig:trust}e, Supplementary Table~1). The insurance records order 75\% of the same pairs. The counties in the top third of predicted loss within a season lost 48\% of their production and those in the bottom third 15\%. Against the university's post-storm county reports the model orders 69\% of 5{,}536 pairs within a storm correctly (64 to 73\%), from 58\% for Idalia to 76\% for Milton.

The probability of a payout is calibrated. In the 1{,}660 storm county-months with $H\ge1$ the Brier score was 0.177 against 0.224 for the historical frequency (0.175 against 0.227 when every parameter is refitted without the year scored), and in ten equal groups of predicted probability the observed proportion with a payout stayed within 8 percentage points of the prediction (Fig.~\ref{fig:trust}f).

The loss checks above are for hurricanes. The freeze and dry-heat percentages are also checked, but less well, because high-quality public data are scarce. For freezes the percentages were compared with state-level and multi-county estimates for the freezes of 2010 and 2026 \cite{roka2010freeze,fdacs2026freeze}; the model gave the right order of magnitude, with extreme events a factor of two to five below the reports and a loss near zero for sugarcane, which the assistant reports when the tier is high. For dry heat the records outside the insurance data were weaker still, because county yields of Florida field crops track the insured-loss label itself only loosely. For vegetables, which are rarely insured in Florida, the assistant marks the percentage as an extrapolation from insured growers.

Estimates for the days ahead rest on forecast weather, so we computed each index and loss from the forecast and from the weather later observed. Up to two days ahead for hurricanes, three for freezes and seven for dry heat, the forecast issued the elevated tier 0.97 to 1.08 times as often as the observed inputs did, reached at least five in six of the severe events, and gave a total loss of 0.90 to 1.23 times that from observed inputs (Methods).

\section{Discussion}\label{sec:discussion}

We asked whether a weather hazard can be carried from public archives to the loss of each crop group on a given day, with every step checked against records. In the demonstration in Florida it can, for hurricanes, freezes and dry heat, and the checks of Results set the level at which the answer holds. Nothing in the procedure is specific to Florida. What a user receives is the tier of the hazard and the expected loss of each crop group as a percentage of its expected season value, and these two numbers are what a decision needs: which crops to protect and which counties to inspect first.

Earlier work approaches the question from three sides, and each stops short of it. Conversational assistants such as ClimSight reason from retrieved text on climate and soil to advise on crop suitability \cite{kuznetsov2025climsight}; they do not run a fitted model. Weather-only indices describe the hazard: the Hurricane Composite Intensity Index behind the UF/IFAS estimates adds the level numbers of wind, rain and flood after the storm and has no published test against recorded losses \cite{court2025idalia,court2025milton}, and the compound index of Chen et al.\ \cite{chen2019ipwt}, the Taiwan loss tabulations \cite{lai2017taiwan} and the indices that rank cyclones for property \cite{holland2019cdp} score an event, not a county on a day or from a forecast. Loss models fitted to insurance records take one hazard with one or a few crops and report a coefficient or a prediction by year, season or month, and none recovers the damage fraction from the payout or checks a hazard-specific loss against production records of the same events \cite{perry2020insurance,biram2025hurricane,reyes2019croploss,seamon2022forest,lazin2025flood,yao2026drought}. WxCrop joins the three: compound indices for every county and day, chosen by causal analysis among physically plausible interacting factors; a loss model with a susceptibility of each crop group to each hazard, giving the loss of a given day from the index and the growth stage of that day, fitted to the damage fraction recovered from the payouts and checked against production records; and an assistant that operates these models and relays their results with their sources. It is the first assistant to carry a weather hazard through fitted crop-loss models to the parcels beneath it.

Because the layers are connected and each stands on its own evidence, one answer serves users at the depth each needs: a specialist can open any layer and find the sourced records behind it, with the assistant assembling them on request, and a grower, an emergency manager or an agency can take the tier and the percentages as directional guidance, which crops to protect first, which counties to send inspectors and relief to first, without reading the layers beneath.

For a hazard still ahead, the same indices run on seven-day forecast inputs and are updated with each new forecast. Several uses follow from this design. Agencies can locate fields for verification and order relief from the tier and crop groups of each county, growers can compare scenarios under a forecast storm, and insurers can anticipate payouts by crop group. In each case the percentages serve as a reference that field verification then refines.

Absolute dollar loss is the hardest output, because no public record states it at a level that could serve as a label: insurance files record payouts after deductibles, post-storm reports give preliminary county estimates and disaster programmes record payments, so these records support percentages far better than dollar figures. A dollar conversion is nevertheless provided in the code repository (Code availability) and checked there against recorded payouts: for the 215 county-months with a recorded payout above US\$100,000 it fell within a factor of ten of the payout in 92\% of cases and within a factor of five in 80\%, and the same check gave 94\% and 79\% for the 159 county-months of freezes and 98\% and 87\% for the 201 county-seasons of dry heat (Supplementary Table~2), an order-of-magnitude reference. The percentage of expected season value remains the primary answer, because it is checked against the records it was fitted to, whereas a dollar figure adds exposure data and prices whose errors compound.

The procedure applies wherever a weather archive, a crop map and a loss record exist, and Florida with its three hazards is one instance of it. A difference-in-differences check on insured losses in nine south-eastern states showed no trend before the storm, a rise of $+0.078$ (95\% interval 0.057 to 0.099) in the damage fraction above the wind line (Methods), and a response that rises with wind, rain and their interaction, with the same behaviour in Florida, Georgia, North Carolina, South Carolina and Alabama. The three families demonstrated here account for 64\% of weather-caused insured payouts in Florida and 69\% across the nine states (Supplementary Table~3), and each has a weather signal that public archives provide for any date. Excess moisture, with 30\% of Florida payouts and 20\% across the nine states, is the next family, and hail and non-cyclone wind follow it. Each needs labels finer than the county-month, such as adjuster reports or field-verified damage, which other states or agencies may hold. The strength of the models is set by the labels that public data provide: insured losses by county, crop group, month and cause, and scarcer records outside the insurance data, so the checks here reach direction and county-scale totals. Farm-level insurance records, adjuster reports or field-verified damage, under data agreements with agencies and insurers, would let the same procedure fit finer than the county, and a state that holds county loss reports for freezes or field-level flood damage can fit and check the models more closely there than Florida allows. Everything the Florida demonstration needed is public, and the path from a public archive to a checked loss estimate, once built, is the same for the next hazard and the next region.
\section{Methods}\label{sec:methods}

The workflow of WxCrop applies anywhere in the world. We demonstrate it on Florida, and the methods below carry to any other region and period by replacing the data sources with those of that region and period and refitting the model to them. Fig.~\ref{fig:overview}b summarises the procedure: four questions settle each hazard index, and the loss model is fitted to the insured losses and checked on years left out and on records outside the insurance data.

\subsection{Data preparation}
The data are of three kinds: weather, agricultural and loss records, the last comprising the insurance payouts that label the loss model and records of the same events outside the insurance data that check it. For each kind we describe the sources, how they were combined and processed, and what a dataset must provide to replace ours when the workflow is extended to another region or period. All three cover the 67 counties of Florida for 2004--2024, the period for which the insured-loss records, the hurricane archive and the gridded weather archives overlap. The hazard index is computed for each county and day, and the loss model is fitted on county $\times$ crop-group $\times$ month units. The month is the unit of the RMA labels and fits the span over which a hazard acts on a crop, and the eight crop groups (citrus, vegetables, other fruit, sugarcane, cotton, peanuts, grains and oilseeds, tobacco) aggregate the 32 insured commodities present in Florida.

\textit{Weather.} Each hazard draws on the archive that records it best (Table~\ref{tab:data}). Tropical-cyclone winds come from HURDAT2, the only record that resolves the core of a cyclone.
\begin{table}[tbp]
  \caption{\textbf{Data used in this study.} Each row gives the quantity, the dataset that supplies it, what it provides and the period used.}
  \label{tab:data}
  \centering
  \fontsize{7}{8}\selectfont
  \renewcommand{\arraystretch}{0.95}
  \setlength{\tabcolsep}{3pt}
  \begin{tabular}{>{\raggedright\arraybackslash}p{0.085\linewidth}>{\raggedright\arraybackslash}p{0.145\linewidth}>{\raggedright\arraybackslash}p{0.32\linewidth}>{\raggedright\arraybackslash}p{0.21\linewidth}>{\raggedright\arraybackslash}p{0.14\linewidth}}
    \toprule
    Kind & Quantity or use & Dataset & Provides & Period \\
    \midrule
    Weather & Wind in cyclones & HURDAT2 best track \cite{landsea2013hurdat2} & Centre, maximum wind and radii, six-hourly & 2004--2024 \\
    \addlinespace[0.5pt]
     & Wind elsewhere & ERA5 \cite{hersbach2020era5}, High-Resolution Rapid Refresh (HRRR) \cite{dowell2022hrrr} & Daily maximum 10-m gust & 2004--2024, HRRR from 2014 \\
    \addlinespace[0.5pt]
     & Threatening cyclone & National Hurricane Center forecast & Forecast track and wind radii & Up to five days \\
    \addlinespace[0.5pt]
     & Rain & Analysis of Record for Calibration (AORC) \cite{fall2023aorc} & Daily precipitation (1 km) & 2004--2024 \\
    \addlinespace[0.5pt]
     & Ponded water, wet hours (tested, not deployed) & National Water Model retrospective \cite{noaaNwmRetro} & Hourly ponded depth and soil saturation (1 km) & 2004--2024 \\
    \addlinespace[0.5pt]
     & Freeze & Parameter-elevation Regressions on Independent Slopes Model (PRISM) \cite{daly2008prism}, ERA5 & Minimum temperature, night cloud cover & Winters 2003--26 \\
    \addlinespace[0.5pt]
     & Drought and heat & gridded Surface Meteorological dataset (gridMET) \cite{abatzoglou2013gridmet} & Precipitation, evapotranspiration, maximum temperature & 2003--2026 \\
    \addlinespace[0.5pt]
     & Missing days, forecast & National Blend of Models \cite{hamill2017nbm} & Gusts, rain, temperature, cloud cover & Forecast days, up to seven \\
    \addlinespace[0.5pt]
     \midrule
    Crops & Which crop grows where & FSAID \cite{fdacsFsaid} & Parcels with crop class and specific crop & Four versions, 2018--2024 \\
    \addlinespace[0.5pt]
     & Crop label check & Cropland Data Layer, Crop Sequence Boundaries \cite{usdaCDL,usdaCSB} & Crop of each pixel or field in each year & Year by year \\
    \addlinespace[0.5pt]
     & When it is in the field & Growing calendars \cite{ifasVH021,nassASB2024,nassAH507,nassAH628} & Planting dates, days to maturity, harvest months & Calendar of a typical year \\
    \addlinespace[0.5pt]
     & What value is at stake & Enterprise budgets of the University of Florida, the University of Georgia and Auburn University \cite{ifasBudgets,ugaBudgets,auburnBudgets} & Cost fraction invested at planting & Budgets of 2001 to 2026 \\
    \addlinespace[0.5pt]
     &  & 2022 Census of Agriculture \cite{usdaCensus2022}, Florida Citrus Statistics \cite{nassCitrusStats} & County crop values & 2022 census, citrus 2023--24 \\
    \addlinespace[0.5pt]
     &  & Tree age and decline status, a screening layer derived here from Landsat composites and NAIP imagery & Age of citrus trees, which sets their value & Parcels with estimates \\
    \addlinespace[0.5pt]
    \midrule
    Losses & Fitting & USDA Risk Management Agency loss records \cite{rmaSobCol} & Payouts and liabilities by county, crop, month, cause & 2004--2025 \\
    \addlinespace[0.5pt]
     & Check, events & NCEI Storm Events Database \cite{nceiStormEvents} & Tropical-cyclone and freeze days & 2004--2024 \\
    \addlinespace[0.5pt]
     & Check, drought & U.S. Drought Monitor \cite{svoboda2002usdm} & Weekly county drought categories & 2000--2026 \\
    \addlinespace[0.5pt]
     & Check, hurricane losses & UF/IFAS post-storm reports \cite{court2023ian,court2025idalia,court2025debby,court2025helene,court2025milton} & County loss ratios & Five storms, 2022--2024 \\
    \addlinespace[0.5pt]
     & Check, production & NASS county production and yields \cite{nassQuickStats,nassCitrusStats} & County production and yield & 1998--2025 \\
    \addlinespace[0.5pt]
     & Check, freezes & UF/IFAS and FDACS freeze reports \cite{roka2010freeze,fdacs2026freeze} & State and multi-county loss estimates & Freezes of 2010 and 2026 \\
    \addlinespace[0.5pt]
    \bottomrule
  \end{tabular}
  \par\smallskip
  {\fontsize{7}{8}\selectfont\noindent Abbreviations: ERA5, fifth-generation ECMWF reanalysis. HURDAT2, Atlantic hurricane database, second generation. FSAID, Florida Statewide Agricultural Irrigation Demand. NCEI, National Centers for Environmental Information. NASS, National Agricultural Statistics Service. UF/IFAS, University of Florida Institute of Food and Agricultural Sciences.\par}
\end{table}

The sources are combined by what each can represent. The wind field of a cyclone is rebuilt from the best-track fixes with a radial profile through the reported wind radii \cite{willoughby2004rmw}, because reanalyses and forecasts smooth the core of a cyclone and underestimate its winds, and the county value is the 90th percentile of grid-cell maxima, which represents the part of the county the core passed over rather than the county mean. On days without a cyclone within 600\,km, wind is the daily maximum gust divided by 1.3 to approximate sustained wind. Rainfall is the county mean of daily totals, log-transformed and standardised, and a standardised precipitation--evapotranspiration index (SPEI) \cite{vicente2010spei} is computed for each county and calendar month. Where an archive has not yet published a day, rainfall falls back to gridMET, and a day that no archive covers takes the forecast inputs described next. On the map the colour boundaries of each layer are percentiles of its own history, from the median to the 99th.

\textit{Forecast inputs.} For the seven days ahead, and for recent days that no archive has yet published, the indices take forecast inputs chosen by the same rule as the observed ones, the source that represents each hazard best (Supplementary Table~4). The National Blend of Models, the blend from here on, supplies them by default, with minimum temperature corrected against PRISM for its warm bias on cold nights and evapotranspiration set to the 2004--2024 mean for the time of year, and a dedicated forecast replaces it where one represents the hazard better. The wind of an Atlantic cyclone that threatens Florida thus comes from the official forecast of the National Hurricane Center for the five days it covers, drawn with the same radial profile as the best track, because the gusts of the blend are too weak at the core of a cyclone (11 to 13\,m\,s$^{-1}$ from the best-track county winds over the storms of 2021--2024, against 4 to 6\,m\,s$^{-1}$ for the official forecast).

We checked the forecast inputs against the observed ones by computing each index and the loss fraction $S_g$ (Equation~\ref{eq:loss}) from both, on every storm that reached Florida in 2004--2024, every winter night of 2020--21 to 2025--26 and every fifth day of April to August in 2021--2025 (Supplementary Table~4). The hurricane check uses the observed rain, and on the 12 storms since the blend was archived in November 2020 the forecast wind and rain together gave 0.84 of the loss from observed inputs.

\textit{Agricultural data.} The data answer three questions, which crop grows where, when it is in the field, and what value is at stake (Table~\ref{tab:data}). The FSAID inventory gives each agricultural parcel a crop class and a specific crop, in four versions with base years from 2018 to 2024, and the version for a date is the latest base year before the target year, or the earliest version for earlier dates.

Growth stage comes from a growing calendar by crop and region (north, central and south Florida), with planting dates and days to maturity for annual crops and harvest months for perennial crops. The vulnerability $v$ of a parcel on the selected date is the fraction of the crop's expected season value at risk, $v=f+(1-f)p$, where $f$ is the fixed fraction held in the plant, the value of the trees for perennial crops and zero for annual crops, and $p$ is the progress of the season on that date, the fraction of the days from planting (or bloom, for fruit crops) to maturity that have passed. For annual crops $p=\max(t/T,\,q)$, with $t$ the days since planting, $T$ the days to maturity and $q$ the establishment cost already invested on the day of planting as a fraction of the season's value, for example the cost of seeds.

\textit{Loss records.} The label of the loss model is the payout, in dollars, recorded by the USDA Risk Management Agency (Table~\ref{tab:data}) by county, commodity and insurance plan. The payout, formally recorded as indemnity, is smaller than the loss in the field because a policy leaves a deductible, a fraction of the crop's expected value, to the grower and pays only the loss beyond it. RMA also provides other necessary complementary data. The first is the liability, the most a policy can pay, which is the product of the expected value in the field in dollars and the coverage level, where the coverage level is one minus the deductible, given for each deductible that the policies of a unit hold. The second is the cause of loss, of which the weather causes are hurricane, tropical storm and tropical depression, excess moisture, wind and flood, and the hurricane subset is used where stated. The third is the mapping rule of how payouts, recorded by month, relate to liabilities, recorded by harvest year. The panel holds 60{,}135 units in 50 counties, each a county, commodity, insurance plan and month of loss. Two further records are prepared alongside the loss records (Table~\ref{tab:data}). One is a record of events that contains no crop information, such as an archive of reported storms, freezes and droughts by county and date. The other is a record of the losses or production of the same events outside the insurance data, from surveys or statistics rather than from the insurance records. Both are used in building the hazard indices below.

\textit{Inversion from payout to damage fraction.} RMA payouts could serve as a good fitting target for the loss model, but for broader value and the convenience of users we recover the damage fraction $d$, the fraction of the crop's expected value lost in the field, and use it as the main outcome of WxCrop. RMA states the policy by which payouts are calculated from $d$, but predicting $d$ in the reverse direction is inexact, for reasons such as a zero payout not meaning zero loss in the field. We use bisection to solve this reverse direction from payout to $d$. For a unit whose liability is spread over deductibles $\delta_k$ in shares $w_k$, a damage fraction $d$ shared by all its insured farms gives the loss cost $\sum_k w_k\max\{0,d-\delta_k\}/(1-\delta_k)$, which rises with $d$, so bisection finds the single $d$ that reproduces the recorded loss cost. A unit with no payout takes $d=0$. From here on the damage fraction $d$ is the label on which the loss model is built. To test whether this inversion method is acceptable, we further need to prepare two datasets: (1) years of the loss records held out of the fit, on which the model is tested for reproducing the damage fraction $d$ on data it has not seen, and (2) a record that does not pass through the insurance files, which chooses among the three candidate labels of Table~\ref{tab:loss}, for which we use the county citrus production of the Florida Citrus Statistics (Loss model).

\textit{Extension to another region.} A replacement weather dataset must be daily, because the index is computed each day and the loss model takes its largest value over 30 days, and gridded finer than the county, because a county mean smooths away the core of a storm and the cold-air pools. It must also be continuous over the period of the loss records and equipped with a forecast counterpart. Where no archive this fine spans the whole period, the next-finest one fills the gap, as ERA5 does for gusts before the HRRR record begins. A replacement agricultural dataset must say which crop grows where, when it is in the field and what value is at stake: a parcel-scale crop map whose crop names link to a calendar, a regional calendar of planting dates and days to maturity or of harvest months, and a value for the crop. A replacement loss dataset must give losses by county, crop and month, separated by cause, with the deductibles of the policies, over enough years to contain many events, and its checks must come from a process unrelated to the loss records, such as post-storm county loss estimates or production statistics.

\subsection{Hazard indices}
We consider four questions in turn while building a hazard index (Fig.~\ref{fig:overview}b): (1) which weather quantities are effective inputs to the index, (2) what basic shape, or family of formula, the index should take, (3) how to make sure that the index marks real damage to crops, and (4) which further factors, weather or site, join the index because they change the loss it brings. We take the hurricane index as the worked example of all four questions and show how the records in Table~\ref{tab:data} are used to settle the form of the index. The freeze and dry-heat indices follow in the same order. The test results for each candidate and the final decisions are in Table~\ref{tab:H} and Supplementary Tables~5 and 6.

To answer the first question, we listed the candidate weather quantities in advance from the physics of the hazard and from earlier indices, and scored each on the event record by how well it alone separated days with a recorded event from days without one, using the AUC introduced in Results. The quantity that dominates the detection became the main input. This step does not exclude the other candidates: they enter the shape chosen in the second question as multipliers or additive terms, and the fourth question keeps each only if, added beside the main input, it explains loss that the main input alone does not.

For hurricanes the candidates were the wind and rain inherited from the IPWT index \cite{chen2019ipwt}, here from HURDAT2 and AORC, and the ponded water and wet hours that we had added to them, from the National Water Model retrospective \cite{noaaNwmRetro}. Against the county-days with a recorded tropical cyclone in the NCEI Storm Events Database \cite{nceiStormEvents} over 2004--2024, the AUC was 0.980 for wind, 0.871 for rain, 0.838 for ponded water and 0.729 for wet hours. Wind became the main input. The other three each detect cyclones well on their own, but on a cyclone day they rise together with the wind, so their scores here may carry nothing beyond it. We kept or dropped each by further tests of what it explains beside the wind in the second and fourth questions.

To answer the second question, we built the formula of the main input in two layers, a shape and its parameter values. A shape is a family of formulas, such as wind above a threshold raised to a power, taken from the indices of the hazard's own literature or from the physics of the damage. It settles the threshold at which damage starts, the window over which the input is counted, and the power by which damage grows past the threshold. Each was tried either fixed at the value that the literature or the physics gives, its prior, or fitted, and a prior was dropped where the fitted value landed far from it, as tropical-storm force was for the wind threshold. The only requirement on a shape is the one that the loss model and the tests below need: it is zero below a reference level and rises above it. The parameters of a shape were fitted together with the crop susceptibilities of the loss model (next subsection), so that the predicted loss fraction comes closest to the recorded damage fraction $d$. The fitted shape was then scored by the AUC of the first question and, where a loss record outside the insurance data exists, by how well it orders the counties against their recorded losses. Both scores depend only on the ranking of the days or counties. A threshold or a window changes that ranking, so the scores tell those choices apart, but a power does not, so the power is set from physics and checked against the loss records.

For hurricanes the shape is inherited from the literature on hurricane damage, wind above a threshold raised to a power \cite{emanuel2011damage,nordhaus2010hurricanes}, the family of Equation~\ref{eq:hazard}. The prior for the threshold is tropical-storm force, 18\,m\,s$^{-1}$, the wind speed from which a storm is named and warned, and the prior for the power is 2, because the pressure that the wind exerts grows with its square. We compared the threshold $w_0$ fixed at 18\,m\,s$^{-1}$ or fitted, and the power 1, 2 or 3, with the window fixed at the 30-day maximum of the daily value, the window of the Data subsection. Each shape was scored on the two records of Table~\ref{tab:data}: by the AUC on the county-days of the first question, and by the share of county pairs that it places in the same order as the county loss ratios of the UF/IFAS post-storm reports, the reported county crop loss over the county's crop sales, the county-order score (Table~\ref{tab:H}). The fitted threshold won the AUC. The powers of 2 and 3 tied on both scores and the power of 1 lost, so we took the power of 2, the dynamic pressure that the wind exerts, and the loss records bore it out: in years left out of the fit (leave-one-year-out, defined in Loss model) it agrees with the insured losses slightly better than the power of 3, which under-predicts the weakest wind class. The index is
\begin{equation}
H=\Big(\frac{(w-w_0)_+}{15}\Big)^{2}\exp(a\,\tilde r),\qquad w_0=4.73\ \mathrm{m\,s^{-1}},\ a=0.221,
\label{eq:hazard}
\end{equation}
where $w$ is the county sustained wind and $\tilde r$ the log rainfall of the county-month standardised against all Florida county-months of 2004--2024, so that $\tilde r=0$ is the historical average rain and the multiplier equals 1 there. Rain is the multiplier adopted in the fourth question, and $w_0$ and $a$ were fitted together once it was adopted. The divisor sets $H=1$ at roughly 20\,m\,s$^{-1}$ at the historical average rain.

To answer the third question, we compared the losses on the two sides of a line drawn through the index by a difference-in-differences (DiD) design with two-way fixed effects \cite{angrist2009mostly,bertrand2004did,goodman2021did,roth2023did}, a design that Biram and colleagues applied previously to the insured losses of crops after hurricanes \cite{biram2025hurricane}. The line is a tier boundary of the index: county-months at or above it count as under the hazard, those below as little or not affected. It is the second boundary of the map legend, where the legend begins to call loss likely, so that the comparison falls on a line that users see, and the third of the four checks below asks whether the loss rises from tier to tier across the remaining boundaries. The loss of a county and crop group above the line is compared with its own loss in months below the line, and that difference with the same difference, in the same month, for the counties that stayed below. What remains, the difference of the two differences, is the effect of crossing the line. Cutting on the strength of the hazard rather than on whether a storm passed keeps the county-months that a storm only brushed below the line, where otherwise they dilute the effect \cite{biram2025hurricane}. The regression is
\begin{equation}
d_{cgt}=\alpha_{cg}+\lambda_t+\beta\,T_{cgt}+\varepsilon_{cgt},
\label{eq:did}
\end{equation}
Two terms are fixed by data, the damage fraction $d$ (Data preparation) of county $c$, crop group $g$ and month $t$, weighted by liability and pooled over crop groups because this step needs only a directional signal of loss, and $T$, which is 1 in the months in which the index is at or above the line and 0 otherwise. The other three are fitted together: $\alpha_{cg}$ is the baseline of the first comparison, everything that stays constant for a county and crop group, such as its usual loss rate, $\lambda_t$ is the common part of the second, everything that a month shares across the state, such as prices or a statewide weather month, and $\varepsilon$ is what remains. The effect $\beta$ is therefore the loss that the county-months above the line carry beyond their own baseline and beyond what the month brought to every county, and an effect of 0.1 means a damage fraction 10 percentage points higher. The equation is fitted by least squares, weighted by liability, with $\alpha_{cg}$ and $\lambda_t$ removed by subtracting the mean of each column within each county and crop group and within each month, which gives the estimate $\hat\beta$ and a residual for every row. The rows of one year share the same storms, so the variance of $\hat\beta$ is estimated with the residuals clustered by year, 21 clusters, and the 95\% interval is $\hat\beta\pm1.96$ standard errors. Four checks follow, each a refit of Equation~\ref{eq:did} with one term replaced (Table~\ref{tab:did-checks}). The interval of $\beta$ excludes zero, so that crossing the line marks more loss. The loss was not already rising in the three months before the crossing, so that it follows the weather. It rises from tier to tier of the index rather than in one step. With the loss fraction $\hat S_{cgt}$ of the loss model (Equation~\ref{eq:loss}, next subsection) in place of $d$, the effect falls inside the interval of the first check, so that the model reproduces the size of the effect that the design finds without it.
\begin{table}[tbp]
  \caption{\textbf{The four checks of the third question.}
    Each check is Equation~\ref{eq:did} refitted with one term replaced, on the same rows, with the same fixed effects, liability weights and year clusters. $t^{*}_{cg}$ is the month in which the index of county $c$ crossed the line, $D^{j}_{cgt}$ marks the rows whose index falls in tier $j$ ($H$ stands for the index of each hazard), and $\hat S_{cgt}$ is the loss fraction $S_g$ of the loss model (Equation~\ref{eq:loss}) evaluated for the row.}
  \label{tab:did-checks}
  \centering
  \small
  \begin{tabular}{>{\raggedright\arraybackslash}p{0.13\linewidth}>{\raggedright\arraybackslash}p{0.40\linewidth}>{\raggedright\arraybackslash}p{0.20\linewidth}>{\raggedright\arraybackslash}p{0.19\linewidth}}
    \toprule
    Check & Regression & Term replaced & Passes when \\
    \midrule
    1. Interval & $d_{cgt}=\alpha_{cg}+\lambda_t+\beta\,T_{cgt}+\varepsilon_{cgt}$ & None & The 95\% interval of $\beta$ excludes zero \\
    2. Leads and lags & $d_{cgt}=\alpha_{cg}+\lambda_t+\sum_{k=-3}^{3}\beta_k L^{k}_{cgt}+\varepsilon_{cgt}$, with $L^{k}_{cgt}=\mathbf{1}[\,t-t^{*}_{cg}=k\,]$ & $T$ by seven month indicators, of which $L^{0}=T$ & The intervals of $\beta_{-3}$, $\beta_{-2}$ and $\beta_{-1}$ include zero \\
    3. Tiers & $d_{cgt}=\alpha_{cg}+\lambda_t+\sum_{j}\beta_j D^{j}_{cgt}+\varepsilon_{cgt}$, with $D^{j}_{cgt}=\mathbf{1}[\,b_j\le H_{ct}<b_{j+1}\,]$ & $T$ by one indicator per tier, $b_1,b_2,b_3$ the legend boundaries, the rows with the index at zero as reference & $\beta_j$ rises from tier to tier \\
    4. Model-implied & $\hat S_{cgt}=\alpha'_{cg}+\lambda'_t+\beta_{\mathrm{model}}\,T_{cgt}+\varepsilon'_{cgt}$ & $d$ by the model's prediction $\hat S$, $T$ unchanged & $\beta_{\mathrm{model}}$ lies inside the interval of check 1 \\
    \bottomrule
  \end{tabular}
\end{table}

For hurricanes, $d_{cgt}$ was the recorded $d$ averaged with liability weights over the other attributes of the records within each county $c$, crop group $g$ and month $t$, the level at which the loss model works (24{,}053 rows). We set the line at $H=2$, which gives $T=1$ for 760 of the 24{,}053 rows. This boundary also has a physical meaning, because $H=2$ is reached at a county wind of about 25\,m\,s$^{-1}$ with average rain, the 50-knot wind that lies midway between tropical-storm and hurricane force and whose radius HURDAT2 reports, and that wind was the threshold fixed for the design before the index existed. All four checks passed (Table~\ref{tab:H}). Reaching $H\ge2$ raised the damage fraction by $+0.165$ (95\% interval 0.120 to 0.210). With the line on the 25\,m\,s$^{-1}$ wind in place of $H$, a threshold that no loss record enters, the rise was $+0.133$ (0.074 to 0.191) in Florida and $+0.078$ (0.057 to 0.099) over nine south-eastern states. Against the county-months with $H=0$, the rise grew steadily with every tier of the index, $+0.008$, $+0.037$, $+0.104$ and $+0.287$, so the loss follows the value of the index. Refitting the same regression with the model's $\hat S$ in place of the recorded $d$ gave $+0.179$, inside that interval, so the jump across the line that the model produces is as large as the one the design finds in the records.

To answer the fourth question, we took as candidates the quantities carried over from the first question and the site factors that could change the damage, such as soil, drainage and the age of trees. A factor can act in two ways, on the loss by itself and on the size of $\beta$, the extra loss of the county-months above the line, and only the second makes it part of the index. We wrote down before the run whether each factor should make $\beta$ larger or smaller. In the design of the third question the two ways are two terms taken out of the residual $\varepsilon$ of Equation~\ref{eq:did}, together with two terms for the main input:
\begin{equation}
d_{cgt}=\alpha_{cg}+\lambda_t+\beta\,T_{cgt}+\gamma\,z^{m}_{cgt}+\theta\,T_{cgt}z^{m}_{cgt}+\delta\,z^{x}_{cgt}+\eta\,T_{cgt}z^{x}_{cgt}+\varepsilon_{cgt},
\label{eq:did-factor}
\end{equation}
where $z^{x}$ is the standardised factor and $z^{m}$ the standardised main input, and the products $Tz^{m}$ and $Tz^{x}$ are the interaction terms. $\delta$ is the effect of the factor by itself and $\eta$, the coefficient of $Tz^{x}$, its change to $\beta$. $\gamma$ and $\theta$ let $\beta$ vary with the main input above the line. Without them, part of the effect of the main input would land in $\hat\eta$ whenever the factor rises on the same days as the main input, as rain does with wind. For a site factor that does not change over time, $z^{x}$ is constant within a county, so $\delta z^{x}$ is absorbed by $\alpha_{cg}$ and only $\eta$ is estimated. Equation~\ref{eq:did-factor} is fitted once per candidate factor, with $d$ and $T$ fixed by data and unchanged from the third question and only $z^{x}$ exchanged, and its intervals are computed in the same way as for Equation~\ref{eq:did}. A factor joins the index when it passes two tests. On the fitted index, it is tried as a multiplier and as an additive term, the form that keeps the AUC and the agreement with the recorded losses is taken, its coefficient ($a$ in Equation~\ref{eq:hazard} for rain) is fitted together with the loss model, and the factor is kept when that coefficient has the expected sign, exceeds chance, and the agreement in the years left out does not fall. On the design, $\hat\eta$ must have the expected sign and exceed chance. The second test is the stricter, because it rests on the design of the third question and not on the form of the loss model. Chance in both is judged by shuffling the factor 200 times and refitting after each shuffle \cite{good2005permutation}, among the months of each county for a weather factor and among the counties for a site factor, and $p$ is the proportion of shuffles whose coefficient reaches the observed one in absolute value. Which tests each factor passed is in Table~\ref{tab:H} and Supplementary Tables~5 and 6.

For hurricanes, rain, the second most effective input after the wind, was added first. As an additive term it lowered the AUC on cyclone days and overpredicted the 25--33\,m\,s$^{-1}$ class, so it entered as the multiplier of Equation~\ref{eq:hazard}, where it passed both tests (Table~\ref{tab:H}): with $T=1$ for the 664 county-months with county wind of at least 25\,m\,s$^{-1}$, the wind at which $H=2$ is reached at average rain, its interaction with the wind exceeded all 200 shuffles, its multiplier coefficient exceeded the shuffled values, and it raised the agreement with the UF/IFAS county order and with the insured losses in left-out years. Ponded water and wet hours, added in turn as further multipliers beside rain, did not pass: their interactions with the wind did not exceed chance, and no pair or triple of multipliers, nor any product of inputs added as a term, improved on rain alone. The other candidates of Table~\ref{tab:H}, from the water-table proxy to tree age, fell short of chance, acted in the opposite direction to what was expected, or were small and tied to rain. Measured water-table anomalies from USGS and water-management-district wells \cite{usgsNwis,swfwmdWmis} passed both tests, but wells exist in only 13 counties, so they are shown as an observation layer rather than entering $H$. With the inputs settled as wind and rain, a free fit of the power and $w_0$ to the damage fraction returned 1.96 and 5.2\,m\,s$^{-1}$. The fitted power is almost equal to the power of 2 chosen in the second question, and fixing it at 2 left both scores unchanged, so the two routes agree on the form of Equation~\ref{eq:hazard}.
\begin{table}[tbp]
  \caption{\textbf{Hurricane index $H$: tests, candidates and decisions.}
    Effects are differences in the damage fraction [95\% cluster interval], defined in Methods. $p$ is the proportion of 200 within-county permutations that reached the observed value (below 0.005 when none did).}
  \label{tab:H}
  \centering
  \scriptsize
  \begin{tabular}{>{\raggedright\arraybackslash}p{0.16\linewidth}>{\raggedright\arraybackslash}p{0.24\linewidth}>{\raggedright\arraybackslash}p{0.33\linewidth}>{\raggedright\arraybackslash}p{0.15\linewidth}}
    \toprule
    Step & Candidate (data) & Result & Decision \\
    \midrule
    Inputs (area under the curve, AUC) & Wind, rain, ponding depth, wet hours (HURDAT2, AORC, National Water Model) & AUC for cyclone days: wind 0.980 (0.946 with one wind source on every day, where $H$ gives 0.955), rain 0.871, ponding 0.838, wet hours 0.729 & Wind main input. Rain, ponding and wet hours to the Factor step \\
    Form (AUC, UF/IFAS agreement, insured losses in left-out years) & Wind above a threshold $w_0$, fixed at 18\,m\,s$^{-1}$ or fitted, exponent 1, 2 or 3 & Fitted $w_0$: exponent 2 0.978, 0.689, 0.361 (ratio 1.02). Exponent 3 0.980, 0.689, 0.356 (ratio 0.98, 18--25\,m\,s$^{-1}$ class 0.85). Exponent 1 0.871, 0.689, 0.379. Fixed $w_0$ 0.864 & Dynamic pressure, fitted $w_0$ \\
    Design (difference-in-differences, DiD) & $H\ge2$, Florida, 24{,}053 rows & Effect $+0.165$ [0.120, 0.210]. Before the storm $-0.016$, $-0.002$, $+0.009$. Tiers $+0.008$, $+0.037$, $+0.104$, $+0.287$. Model $+0.179$ & Passed \\
    Factor (permutation) & Rain $\times$ wind (AORC) & Amplifying wind $+0.066$ [0.030, 0.102], $p<0.005$. Multiplier 0.22 against 0.096. UF/IFAS 0.689 to 0.704, left-out years 0.361 to 0.397 & Adopted, multiplier \\
     & Ponding, wet hours (National Water Model) & Amplifying wind $+0.024$ ($p=0.35$), $-0.023$ ($p=0.80$). As multipliers beside rain, left-out years 0.381, 0.378. Alone $+0.017$, $+0.006$, $p<0.005$, small and tied to rain & Set aside \\
     & Water-table depth, drainage class, tree age, days since storm & $p=0.27$--0.98, or sign opposite & Set aside \\
     & Wind duration (HURDAT2 hourly field) & $+0.002$, $-0.014$ with peak wind controlled & Set aside \\
     & Measured water-table anomaly (wells) & Alone $+0.008$ [0.002, 0.014], $p<0.005$. Amplifying wind $\hat\eta=+0.068$ [0.015, 0.121], $p=0.015$. Wells in 13 counties & Observation layer \\
    \bottomrule
  \end{tabular}
\end{table}

\textit{Freeze index.} Following the same methodology, with the candidates, scores and decisions in Supplementary Table~5, we derived the formula of the freeze index,
\begin{equation}
F_z=\Big(\frac{1}{N}\sum_{k=1}^{N}\big(T_{d,k}-T_{\min,k}\big)_+\Big)\exp\!\Big(\frac{c\,z_{\mathrm{cloud}}}{2}\Big),\qquad c=-0.24,
\label{eq:freeze}
\end{equation}
where $T_{\min,k}$ is the daily minimum temperature of farmland cell $k$ on that night, from PRISM, over the $N$ cells of the county, $T_{d,k}$ the historical baseline of the cell, so that the index measures how far the night goes beyond the normal state of the place, as $H$ does for rain, and $z_{\mathrm{cloud}}$ the standardised night cloud cover. Because the quantity is a minimum, $T_{d,k}$ is set at the fifth percentile of $T_{\min,k}$ over the winter nights of 2004--2024 rather than at the coldest night on record, so that a single extreme does not draw the baseline to a very low value. The percentile was chosen in the second question, where it detected the freeze days of the Storm Events Database best. Of the further factors of Supplementary Table~5, cloud cover alone passed the fourth question with information that the depth below $T_{d,k}$ does not already carry, and the form agrees with the physics of radiation frost: on a clear night the canopy loses heat by long-wave radiation to the sky and falls below the air temperature, whereas cloud returns part of that radiation \cite{snyder2005frost}, so $c<0$ and the same cold gives a larger $F_z$ on a clear night.

\textit{Dry-heat index.} The dry-heat index came out of the same four questions, recorded in Supplementary Table~6, as
\begin{equation}
D_z=\big(\mathrm{SPEI}_{120}^{+}+0.80\,\mathrm{SPEI}_{30}^{+}\big)\exp(0.117\,z_{\mathrm{heat}}),
\label{eq:dry}
\end{equation}
where $\mathrm{SPEI}_k$ is the $k$-day deficit of precipitation minus reference evapotranspiration, standardised against the same calendar month of the county over 2004--2024 \cite{vicente2010spei}, $\mathrm{SPEI}_k^{+}$ its positive part, which keeps the dry side, and $z_{\mathrm{heat}}$ the standardised 30-day sum of degree-days above 32\,$^\circ$C, so that the index measures how far the season is drier and hotter than the normal state of the place. The unit of time in Equation~\ref{eq:did} was the April--September season of each county and crop group, because drought losses are recorded at harvest rather than in the month of the deficit. The first check of the third question did not pass, so the index serves the tiers and the ranking of county-seasons of field crops rather than a mean effect. Of the further factors of Supplementary Table~6, the short deficit and heat alone passed the fourth question, and the form agrees with the physics: a 30-day deficit catches the rapid-onset drought that the 120-day deficit misses \cite{otkin2018flash}, so it adds to the deficit, and heat raises the evaporative demand and the stress of the crop at the same deficit \cite{lesk2016extremes}, so it multiplies it. The index applies to field crops, with irrigated citrus, vegetables and sugarcane responding little. Excess moisture outside tropical cyclones was examined with a heavy-rain and wet-spell index against recorded flood days and insured claims, and only its rain tier was retained.

\subsection{Loss model}
The hazard indices contain weather only, so that they have a value for every day and mean the same for every crop. The loss model turns the index into the damage fraction by adding the susceptibility $\theta_g$ of each crop group $g$ to each hazard and the stage of the season (Fig.~\ref{fig:overview}b, fit and checks). It was built by the same four questions. The first two are those of the index, its input. The third becomes which crop groups the hazard reaches and how large its effect is for each, and the fourth whether any further weather or site variable changes the effect beyond the index.

The loss model is fitted to $d$, and its accuracy is judged on years it has not seen: each of the 21 years is left out in turn, the model is fitted on the others, and the year left out is predicted. The Spearman rank correlation \cite{spearman1904rank} between the predicted and the recorded damage is taken across the storm county-months of those predictions, those with county wind of at least 18\,m\,s$^{-1}$, and measures how well the counties are ordered by loss, and totals are compared as the ratio of predicted to recorded sums. The score and the questions answer different things. The design of Equation~\ref{eq:did} says whether the hazard changes the recorded loss and which factors change that effect, without any model, but it cannot score a form. The score says whether a form predicts the years it was not fitted on, but it cannot tell a cause from an association. A term therefore enters the model only when it passes the questions, and a form is kept only when it scores.

The hurricane loss model is again the worked example, and the other two hazards are noted where they differ. For the third question, a direction was all that was needed for the hazard index, so one pooled $\hat\beta$ over all crop groups served. For the loss, Equation~\ref{eq:did} is fitted by the same method for each crop group on its own, with a $\hat\beta_g$ of its own for the same line $H\ge2$ (Table~\ref{tab:loss}). The groups run from cotton and citrus at the top to other fruit and sugarcane, whose intervals contain zero and whose $\theta_g$ is expected to be near zero as well. The $\theta_g$ are then fitted, and the fourth check of the third question is repeated by group: with $\hat S$ in place of $d$ the same regression gives the rise the model implies for the group, and where it lies inside the design interval of $\hat\beta_g$ the size the model assigns to the group is supported by the data rather than by the formula. It does so for seven of the eight groups. For vegetables the model implies about twice the recorded rise, possibly because vegetable policies are mostly dollar plans with a low insured share, so that the record understates their losses, or because $\theta_g$ runs high there. The steps are in Table~\ref{tab:loss}.
\begin{table}[tbp]
  \caption{\textbf{Loss model: tests, candidates and decisions.}
    Rank correlations are Spearman rank correlations. Effects of $H\ge2$ are design estimates on the damage fraction by crop group.}
  \label{tab:loss}
  \centering
  \scriptsize
  \begin{tabular}{>{\raggedright\arraybackslash}p{0.16\linewidth}>{\raggedright\arraybackslash}p{0.24\linewidth}>{\raggedright\arraybackslash}p{0.33\linewidth}>{\raggedright\arraybackslash}p{0.15\linewidth}}
    \toprule
    Step & Candidate (data) & Result & Decision \\
    \midrule
    Label & Damaged acreage, loss cost, recovered damage fraction & Rank correlation with county citrus production shortfalls across 72 county-seasons 0.36, 0.47, 0.58 & Recovered damage fraction \\
    Effect by crop group & $\hat\beta_g$ of Equation~\ref{eq:did} fitted for each group, line $H\ge2$ & Citrus $+0.30$, cotton $+0.32$, peanuts $+0.16$, tobacco $+0.13$, vegetables $+0.06$, grains and oilseeds $+0.04$, other fruit $+0.02$, sugarcane $-0.06$. Model-implied effect inside the interval for 7 of 8 groups & Susceptibility $\theta_g$ by group \\
    Structure & IPWT form on the four quantities with equal weights, wind alone, $H$ & Storm-month rank correlation, leave-one-year-out, 0.351, 0.361, 0.397 & $S_g=1-\exp(-\theta_g\varphi_g^{\gamma}H)$ \\
    Curve & $1-e^{-S^{*}_g}$, $S^{*}_g/(1+S^{*}_g)$, $\tanh(S^{*}_g/2)$, all parameters refitted & Leave-one-year-out rank correlation: hurricane 0.397, 0.396, 0.396; freeze 0.334, 0.334, 0.334; dry heat 0.577, 0.580, 0.575 & $1-e^{-S^{*}_g}$, the result insensitive to the curve \\
    Yield vulnerability & Floor of 0.02, establishment-cost fractions; county parcel vulnerability for annual crops & Rank correlation 0.397 against 0.397, totals 1.02 against 1.02; parcel vulnerability 0.41 against 0.40, totals 1.01 against 1.02 & Establishment-cost fractions \\
    Storm periphery, moisture & Fraction of county $\ge18$\,m\,s$^{-1}$, thresholds by group, hours above 18\,m\,s$^{-1}$, rainfall extremes, wet-day runs, soil moisture, ponding & Every metric within the permutation spread & Set aside \\
    Payout probability & $P_g$ with a background rate and a storm scale & Brier score 0.177 against 0.224 for the climatological rate. Four checks with the payout indicator: effect $+0.36$ [0.28, 0.43], before and after within zero, tiers $+0.04$, $+0.14$, $+0.30$, $+0.60$, model-implied $+0.31$ inside, inside for all eight groups & Adopted \\
    \bottomrule
  \end{tabular}
\end{table}

The loss fraction is built from $S^{*}_g=\theta_g\varphi_g(t)^{\gamma}H$: the index, scaled by the susceptibility $\theta_g$ of the group, which the step above showed to differ by crop, and by the yield vulnerability $\varphi_g(t)$ (defined below). The exponent $\gamma$ lets the data decide how strongly the date matters: $\gamma=1$ would make the loss proportional to $\varphi_g(t)$, $\gamma=0$ would remove the date, and the fitted 0.45 lies between, because a young planting can be destroyed outright while little of its yield is in the field. The damage of a crop group cannot be below 0 or above 1, and it rises with $S^{*}_g$ ever more slowly as it nears 1. We chose a form with these three properties,
\begin{equation}
S_g=1-\exp\!\big(-\theta_g\,\varphi_g(t)^{\gamma}H\big),
\label{eq:loss}
\end{equation}
Other forms with the same properties, $S^{*}_g/(1+S^{*}_g)$ and $\tanh(S^{*}_g/2)$, give almost the same rank correlation in years left out with all parameters refitted (Table~\ref{tab:loss}), so the loss is insensitive to the shape of the curve and Equation~\ref{eq:loss} is used. The form returns $S^{*}_g$ itself when it is small, as the additive effect of Equation~\ref{eq:did} does, and against the IPWT form \cite{chen2019ipwt} carried to the four candidate quantities with equal weights, and against wind alone, it ranked the storm months of years left out of the fit best (Table~\ref{tab:loss}).

The yield vulnerability $\varphi_g(t)$ is the vulnerability $v$ of the Data subsection without its fixed fraction $f$, that is, the progress $p$ with its establishment-cost floor $q$, averaged over the crops of the group weighted by insured liability and over regions, so that it lies between 0.02 and 1, reaches 1 at maturity and stays there through the harvest months, with perennial groups at 1 throughout. The fixed fraction $f$ is left out because it lies outside what crop insurance pays. Raising the floor from 0.02 to the establishment-cost fraction, and replacing $\varphi_g$ for annual crops by the acreage-weighted parcel vulnerability of each county, each left the Spearman rank correlation in years left out unchanged (Table~\ref{tab:loss}).

For the fourth question, further terms were sought where the model fell short. Counties at the edge of the wind field were under-predicted, so terms for the periphery of a storm (the fraction of the county at or above 18\,m\,s$^{-1}$, thresholds by crop group, hours above 18\,m\,s$^{-1}$, and rainfall extremes within the county) were tried, and losses in wet months without strong wind suggested terms for moisture (runs of wet days, soil moisture and parcel ponding). Each was declared before it was fitted and tested by the rules of that question, and all were set aside, because none was larger than chance.

The least-squares fit to $d$ over all county-months, those without a payout included, weighted by liability and with a small penalty on the size of $\theta_g$, gives the $\theta_g$ of Supplementary Table~7, from cotton 0.116 to sugarcane 0.001, and $\gamma=0.45$. Because the months without a payout are in the fit, $S_g$ is the expected loss fraction given the index. For a date, $S_g$ is evaluated at the largest daily index over the 30 days that end on it, the window of the Data subsection. $S_g$ and the county values on the map are on the insurance scale, the damage on the insured units of the group, the scale on which the model was fitted. The county value is the mean of the $S_g$ of its crop groups weighted by their expected season value.

The finished model is then scored once, on the predictions for the years left out, and checked against records outside the insurance files. On the insured losses, the Spearman rank correlation and the ratio of totals are those of Table~\ref{tab:loss}, 0.40 and 1.02, on storm county-months and on the insurance scale, and Supplementary Table~8 gives the ratio by wind class and by crop group and the share of the 20 largest recorded county-months predicted within a factor of two. The rank correlation is read against the label itself, which agrees with the county citrus production shortfalls at 0.58 (Table~\ref{tab:loss}), and it ranks every storm county-month, the many small ones included. The insured units of one county, crop group and month, split at random into two halves, agree with each other at 0.59 on the 1,141 storm county-months that hold two or more units. The external records entered no fit. The UF/IFAS loss ratios scored the shapes of $H$ in the second question, the same 0.689 for every exponent (Table~\ref{tab:H}), and the citrus production records chose $d$ among the three candidate labels of Table~\ref{tab:loss}. County citrus production from the Florida Citrus Statistics county tables \cite{nassCitrusStats} enters as the shortfall of each county against its most recent season without hurricane-force wind, net of the county's median change over calm seasons. It is compared by the county-order score with the largest daily citrus $S_g$ from June to November before the harvest, within each of the five seasons in which county wind $w$ reached hurricane force, 33\,m\,s$^{-1}$, in a citrus county, on the counties with production reported in both seasons. The county loss ratios of the UF/IFAS reports are compared by the same score within each storm with the liability-weighted mean over crop groups of the largest daily $S_g$ in the storm window of the report. Pairs that share a county or a storm are dependent, so the 95\% interval of each score resamples the counties, each with all its seasons or storms. The outcomes are in Results and Fig.~\ref{fig:trust}.

$S_g$ is a mean over months with and without a payout, so a second model gives the probability of a payout,
\begin{equation}
P_g=1-(1-b_g)\exp\!\big(-k_g\,S^{*}_g\big),
\label{eq:claim}
\end{equation}
where $b_g$ is the payout rate where $S^{*}_g=0$ and $k_g$ scales the rise with $S^{*}_g$, both fitted by maximum likelihood to whether any weather-cause payout was recorded (Supplementary Table~7). $P_g$ passed the same four checks of the third question with the payout indicator in place of $d$, pooled and for each group, and its calibration in years left out is scored by the Brier score, the mean squared difference between the predicted probability and the outcome (Table~\ref{tab:loss}). It is reported beside $S_g$, at the same 30-day index.

The freeze and dry-heat layers are Equation~\ref{eq:loss} with their own index in place of $H$. For freezes, $S_g=1-\exp(-\theta_g\varphi_g(t)^{\gamma}F_z^{2})$ with $\gamma=2.5$, the square entering as $(w-w_0)_+^{2}$ does in $H$. For dry heat, on the season units of the previous subsection, $S_g=1-\exp(-\theta_g\bar D_z^{\,\rho})$ with $\rho=0.46$, where $\bar D_z$ is $D_z$ averaged over the April--September season with the growing calendars as weights, which take the place of $\varphi_g$. The susceptibilities $\theta_g$ and Equation~\ref{eq:claim} are fitted for each hazard to the insured losses of its own causes (freeze, frost, cold wet weather and cold winter, and drought, heat, hot wind and failure of irrigation supply), with Brier scores in years left out of 0.050 against 0.063 on the exposed winter rows and 0.137 against 0.227, and all fitted parameters in Supplementary Table~7.

\subsection{Assistant}
The assistant is a language model (gpt-6-sol) that reads the conversation and chooses which tools of WxCrop to call and with what arguments. The host runs each call and records a receipt: the expression evaluated, its result, its unit, the version of the data product and its hash. The model answers by selecting receipts and writing the qualitative text, and the host renders every number, unit and filter in the reply from the selected receipts. A number that no receipt of the current turn supports is not delivered, so the figures in a reply are computed by WxCrop and traceable to their records rather than written by the language model.

\subsection{Statistics and reproducibility}
No experiment was performed, and no randomisation or blinding applies: every analysis is on public records, listed with their sources, periods and resolutions in Table~\ref{tab:data} (Data preparation). The difference-in-differences design of each hazard index is Equation~\ref{eq:did}, and its four checks, each a refit with one term replaced, are defined in Table~\ref{tab:did-checks} (Hazard indices, third question). The samples, effects and intervals of the designs are in Table~\ref{tab:H} for hurricanes and in Supplementary Tables~5 and 6 for freezes and dry heat. Intervals are two-sided 95\% intervals with standard errors clustered by year. Permutation $p$ values are defined under the fourth question and reported beside each coefficient in the same tables. The leave-one-year-out scores, the Spearman rank correlation, the ratio of totals and the Brier score, are defined in Loss model and reported in Table~\ref{tab:loss} and Supplementary Tables~7 and 8 for the three hazards. The county-order score and its pair counts are in Results and Loss model. Every number in the paper traces to one script and one result file in the repository (Code availability).

\subsection{Code availability}

The code of WxCrop and of the analyses reported here, and the public web interface of WxCrop, will be made public when the paper is published.

\section*{Data availability}

All source data are public and are listed with their citations in Methods, Data preparation. The scripts that download and process them and the derived tables of every analysis reported here will be released with the code.

\section*{Use of AI tools}
Language-model assistants (Claude, Anthropic, and ChatGPT, OpenAI) were used under the authors' direction to draft and edit the text and to write and check the analysis and platform code. The authors designed the study, verified every analysis, number and citation, and take full responsibility for the content. The assistant inside WxCrop is itself built on a language model, as described in Methods.

\bibliography{refs}

\includepdf[pages=-]{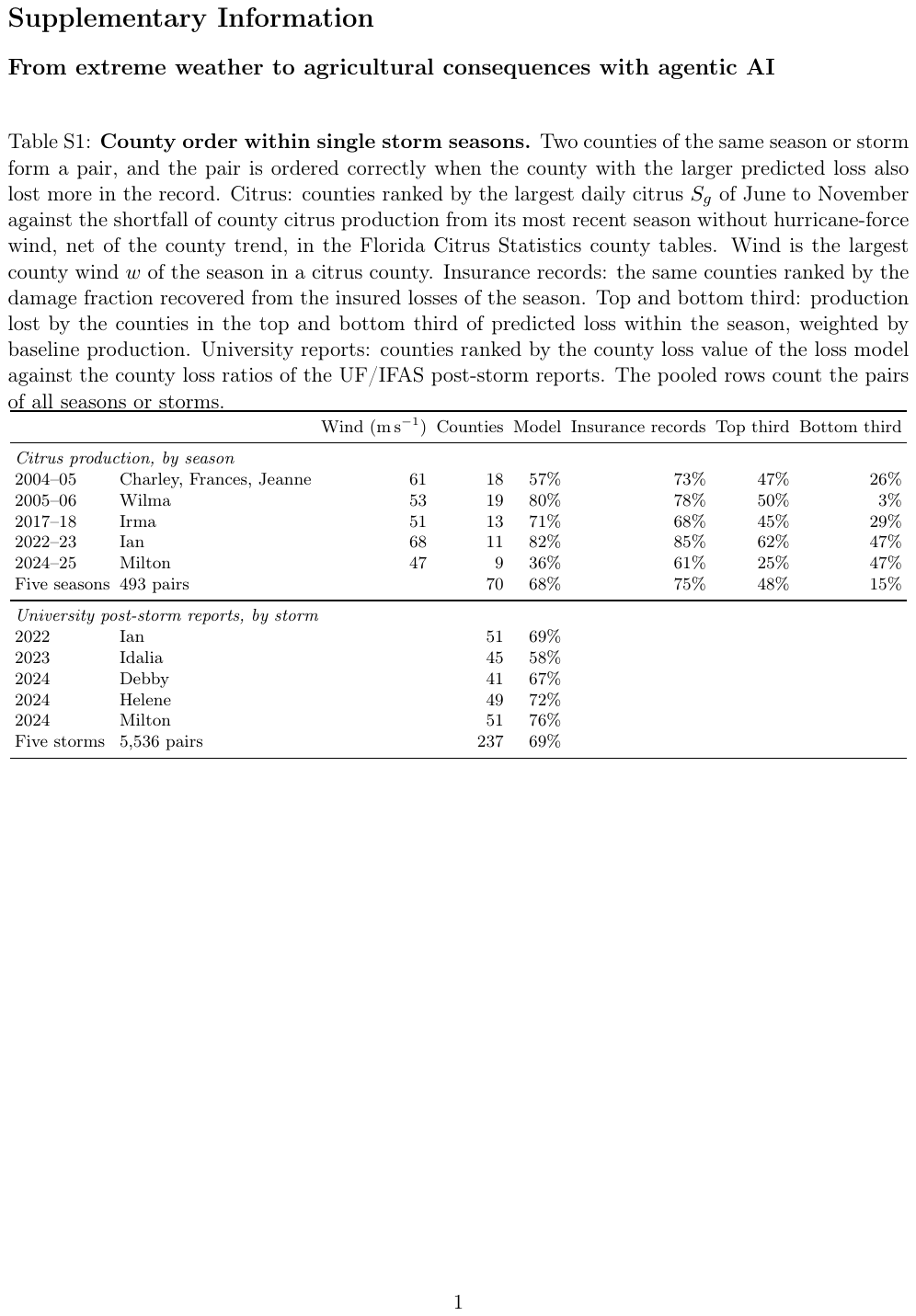}

\end{document}